\documentclass[letterpaper]{aa}

\usepackage{graphicx}
\usepackage{url}

\newcommand{\EQ}{\begin{equation}}
\newcommand{\EN}{\end{equation}}

\newcommand{\Eq}[1]{Eq.~(\ref{#1})}
\newcommand{\rr}{{\vec{r}}}
\newcommand{\uu}{{\vec{u}}}
\newcommand{\BB}{{\vec{B}}}
\newcommand{\JJ}{{\vec{J}}}

\newcommand{\FF}{{\vec{F}}}
\newcommand{\AAA}{{\vec{A}}}
\newcommand{\SSS}{{\vec{S}}}
\newcommand{\ttau}{{\vec{\tau}}}

\newcommand{\dd}{{\rm d} {}}
\newcommand{\DD}{{\rm D} {}}

\newcommand{\km}{\,{\rm km}}
\newcommand{\s}{\,{\rm s}}
\newcommand{\K}{\,{\rm K}}
\newcommand{\G}{\,{\rm G}}
\newcommand{\kG}{\,{\rm kG}}
\newcommand{\AU}{\,{\rm AU}}
\newcommand{\g}{\,{\rm g}}
\newcommand{\cm}{\,{\rm cm}}
\newcommand{\yr}{\,{\rm yr}}

\newcommand{\cs}{{c_{\rm s}}}

\newcommand{\erg}{\,{\rm erg}}

\usepackage{times}

\begin{document}


\titlerunning{Outflows and accretion in a star--disc system}
\title{Outflows and accretion in a star--disc system with\\
       stellar magnetosphere and disc dynamo}

\author{
  B.~von~Rekowski\inst{1}
  \and A.~Brandenburg\inst{2}
}

\institute{
  Department of Astronomy \& Space Physics,
  Uppsala University, Box 515, 751 20 Uppsala, Sweden
  \and NORDITA, Blegdamsvej 17, DK-2100 Copenhagen \O, Denmark
}

\offprints{Brigitta.vonRekowski@astro.uu.se}

\date{Received  / Accepted}


\abstract{
The interaction between a protostellar magnetosphere and a surrounding
dynamo-active accretion disc is investigated using an axisymmetric
mean-field model. In all models investigated, the dynamo-generated
magnetic field in the disc arranges itself such that in the corona,
the field threading the disc
is anti-aligned with the central dipole so that no
{\sf X}-point forms.
When the magnetospheric field is strong enough (stellar surface field strength
around $2\kG$ or larger), accretion happens in a recurrent fashion
with periods of around 15 to 30~days,
which is somewhat longer than the stellar rotation period of around 10~days.
In the case of a stellar surface field strength of at least a few $100\G$,
the star is being spun up by the magnetic torque exerted on the star.
The stellar accretion rates are always reduced by the presence of a
magnetosphere which tends to divert a much larger fraction of the disc
material into the wind.
Both, a pressure-driven stellar wind and a disc wind form. In all our models
with disc dynamo, the disc wind is structured and driven by magneto-centrifugal
as well as pressure forces.
}
\maketitle

\section{Introduction}

\begin{figure}[t!]
\centerline{
   \includegraphics[width=8.5cm]{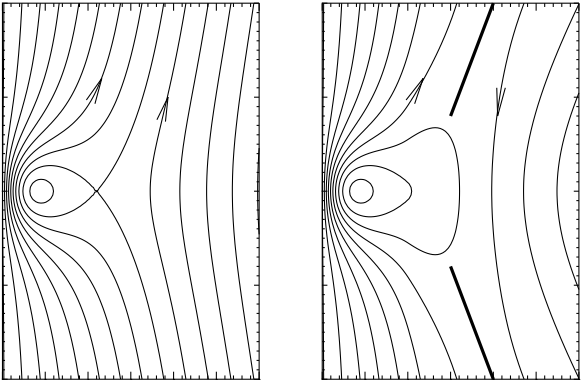}
}
\caption[]{
Sketch showing the formation of an {\sf X}-point when the disc field is
aligned with the dipole (on the left) and the formation of current sheets
with no {\sf X}-point if they are anti-aligned (on the right).
The two current sheets are shown as thick lines.
In the present paper, the second of the two configurations emerges
in all our models,
i.e.\ with current sheets and no {\sf X}-point.
}\label{xpoint}
\end{figure}

The interaction of a stellar magnetic field with a circumstellar
accretion disc was originally studied in connection with accretion discs
around neutron stars (Ghosh et al.\ 1977, Ghosh \& Lamb 1979a,b),
but it was later also applied to protostellar magnetospheres
(K\"onigl 1991, Cameron \& Campbell 1993, Shu et al.\ 1994).
Most of the work is based on the assumption that the field in the
disc is constantly being dragged into the inner parts of the
disc from large radii.
The idea behind this is that a magnetized molecular cloud collapses,
in which case the field in the central star and that in the disc are aligned
(Shu et al.\ 1994).
This was studied numerically by Hirose et al.\ (1997) and Miller \& Stone (1997).
In the configurations they considered, there
is an {\sf X}-point in the equatorial plane
(see left hand panel of Fig.~\ref{xpoint}),
which can lead to a strong funnel flow.

Another alternative has been explored by Lovelace et al.\ (1995)
where the magnetic field of the star has been flipped and is now
anti-parallel with the field in the disc, so that the field in the
equatorial plane points in the same direction and has no {\sf X}-point.
However, a current sheet develops above and below the disc plane
(see right hand panel of Fig.~\ref{xpoint}).
Numerical simulations of such a field configuration
by Hayashi et al.\ (1996) confirm the idea by Lovelace et al.\ (1995)
that closed magnetic loops connecting the star and the disc are twisted
by differential rotation between the star and the disc, and then inflate
to form open stellar and disc field lines (see also Bardou 1999,
Agapitou \& Papaloizou 2000). Goodson et al. (1997,1999)
and Goodson \& Winglee (1999) show that for sufficiently
low resistivity, an accretion process develops that is unsteady and proceeds
in an oscillatory fashion. The inflating magnetosphere expands to
larger radii where matter can be loaded onto the field lines and be ejected as
stellar and disc winds. Reconnection of magnetic field lines allows matter
to flow along them and accrete onto the protostar, in the form of a funnel flow
(see also Romanova et al.\ 2002). Consequently, their stellar jets show
episodic behaviour; see also Matt et al.\ (2002).

The gas in the disc is turbulent and hence capable
of converting part of the kinetic energy into magnetic energy by dynamo action.
Furthermore, the differential rotation of the disc can allow for large scale
magnetic fields, possibly with spatio-temporal order (similar to
the solar 11 year cycle; see, e.g.,\ Parker 1979).
Such a dynamo may have operated in the solar nebula and in
circumstellar or protoplanetary discs
(e.g.,\ Reyes-Ruiz \& Stepinski 1995).
However, not much is known about the mutual interaction between
a stellar magnetic field and a dynamo-generated disc field.
In particular, we need to understand the effect of the magnetic field
generated in the disc, and we also want to know how this is being affected
by the presence of the central dipole field.
Following a recent attempt to model outflows from cool, dynamo-active
accretion discs (von Rekowski et al.\ 2003; hereafter referred to
as Paper~I), we present in this paper a study of the
interaction between the dynamo-generated disc magnetic field and
the field of the central star.
In our model, both
the star and the disc have a wind, resulting in a structured outflow
that is driven by a combination
of different processes including pressure-driving
as well as magneto-centrifugal acceleration.

\section{The magnetospheric model} \label{TMM}

A detailed description of the model without magnetosphere can be found
in Sect.~2 of Paper~I.
We begin the description of the present model by reviewing the basic setup
of the model in Paper~I.
The implementation of the stellar magnetosphere is described in
Sect.~\ref{ADDSM}.

\subsection{Basic equations} \label{BE}

We solve the set of axisymmetric MHD equations
in cylindrical polar coordinates $(\varpi,\varphi,z)$,
consisting of the continuity equation,
\begin{equation}
   {\partial\varrho\over\partial t}+\vec{\nabla}\cdot(\varrho\uu)
   =q_\varrho^{\rm disc}+q_\varrho^{\rm star},
\label{Cont}
\end{equation}
the Navier--Stokes equation,
\begin{equation}
   \frac{\DD\uu}{\DD t}
   = - {1\over\varrho}\vec{\nabla}p - \vec{\nabla}\Phi
     + \frac{1}{\varrho}\left[\vec{F} + (\vec{\uu}_{\rm K}-\vec{\uu})
       q_\varrho^{\rm disc}\right],
\label{Momentum}
\end{equation}
and the mean field induction equation,
\begin{equation}
   {\partial\AAA\over\partial t}=\uu\times\BB+\alpha\BB-\eta\mu_0\JJ.
\label{Ind}
\end{equation}
Here, $\uu$ is the velocity field, $\varrho$ is the gas density,
$p$ is the gas pressure, $\Phi$ is the gravitational potential,
$t$ is time, ${\DD}/{\DD}t=\partial/\partial t+\uu\cdot\vec{\nabla}$
is the advective derivative,
$\FF=\JJ\times\BB+\vec{\nabla}\cdot\ttau$ is the sum of the Lorentz
and viscous forces, $\JJ=\vec{\nabla}\times\BB/\mu_0$ is the current
density due to the mean magnetic field $\BB$, $\mu_0$ is the magnetic
permeability, and $\ttau$ is the (isotropic) viscous stress tensor.
In the disc, we assume a turbulent (Shakura--Sunyaev) viscosity,
$\nu_{\rm t}=\alpha_{\rm SS}c_{\rm s}z_0$,
where $\alpha_{\rm SS}$ is the Shakura--Sunyaev coefficient (less than unity),
$c_{\rm s}=(\gamma p/\varrho)^{1/2}$ is the sound speed, $\gamma = c_p/c_v$
is the ratio of the specific heat at constant pressure, $c_p$, and
the specific heat at constant volume, $c_v$,
and $z_0$ is the disc half-thickness.

As described in Paper~I,
in order to maintain a statistically steady accretion disc, we
need to replenish the mass that is accreted through the disc and
onto the star. We therefore include
a mass source, $q_\varrho^{\rm disc}$, that is restricted to the disc
and self-regulatory, i.e.\ it turns on once the local density in the
disc drops below the initial density distribution, $\varrho_0(\rr)$,
of the hydrostatic equilibrium.
This means that matter is injected into the disc only wherever and whenever
$\varrho<\varrho_0$, and the strength of the mass source is proportional
to the gas density deficit. Therefore, we do not prescribe the distribution
and magnitude of the mass source beforehand, but the system adjusts
itself.
We also allow for a self-regulatory mass sink, $q_\varrho^{\rm star}$,
at the position of the central object (protostar) to model accretion
onto the central star without changing the stellar radius.
The mass sink is modelled in a way analogous to
the mass source.
In the models with magnetosphere, however, a mass sink in the star is modelled
by setting density, velocity and magnetic field in the star to their initial
values at each timestep. This is necessary in order to anchor
the magnetosphere in the star (cf.\ Sect.~\ref{ADDSM}).
The mass source appears also in the Navier--Stokes equation, unless
matter is injected with the ambient velocity of the gas.
We always inject matter with Keplerian speed,
$\vec{\uu}_{\rm K}$. This leads to an extra term in the Navier--Stokes
equation, $(\vec{\uu}_{\rm K}-\vec{\uu}) q_\varrho^{\rm disc}$, which
vanishes only if the gas rotates already with Keplerian speed or when no mass
is injected ($q_\varrho^{\rm disc}=0$).

A physically realistic accretion disc is much more
strongly stratified than what can be represented in the simulations.
We expect most of the actual accretion to occur
in the innermost parts of the disc.
These inner parts are also much cooler and therefore they
should be spinning at almost exactly Keplerian speed, i.e.\
usually faster than the outer parts of the disc. This is the reason why
we choose to inject new matter at Keplerian speed. One additional reason
why we do not inject matter with the ambient gas velocity
is that we want to prevent a runaway effect (cf.\ Sect.~2.1 in Paper~I).
Such a runaway could result in a loss of the initial angular momentum
of the disc, which would never be replenished, so that the entire disc
would eventually lose its angular momentum and all matter would be accreted.
However, in some cases we have compared simulations with and without
Keplerian injection and found the difference to be small.

Furthermore,
mass replenishment in our disc is necessary, because we model an accretion disc
that is truncated at about $0.19\AU$ (cf.\ Sect.~\ref{DVCP}), i.e.\ we model
the inner part of a realistic (protostellar) accretion disc. The total run time
of our simulations is between $\sim$ 150~days (Model~S) and $\sim$ 1900~days
(Model~M1), corresponding to about 33 to 422 Keplerian orbital periods
of the inner disc edge (cf.\ Sects~\ref{DVCP} and \ref{RES}). This is shorter
than the life time of the inner part of a protostellar accretion disc.
On the other hand, a typical advection time in our models is
$(\varpi_{\rm out} - \varpi_{\rm in})/u_\varpi \approx (1.9-0.6)\times0.1\AU/
(0.15\times10^2\km\s^{-1}) \approx$ 15~days (cf.\ Sects~\ref{DVCP} and
\ref{SM}), which is much shorter than the typical run time so that our disc would
disappear if there were no mass supply.
One way to model mass accretion from the outer parts of the disc to the inner
parts, is to extend
our disc to the radial boundary of our computational domain and to inject matter
at this radial boundary. Another way -- that we have chosen -- is to inject
matter locally in the disc wherever and whenever it is needed.

The technique of invoking self-regulatory terms that act in certain
parts of the computational domain is an alternative to prescribing
boundary conditions.
Similar techniques have been used in modelling the Earth's magnetosphere
(Janhunen \& Huuskonen 1993) and in magnetized Couette flow problems
where inner and outer cylinders are present as part of a cartesian mesh
(Dobler et al.\ 2002).
In the present case, our star and accretion disc play a hybrid r\^ole in that
they represent {\it not only} boundary conditions to the corona
including the region between disc and star.
In the models without magnetosphere, the star is not only a region where mass
is being absorbed by the mass sink, but one can also study the dynamics
of the velocity and magnetic fields in the star and the resulting effects.
We do not prescribe inflow at the star's surface; instead, our model allows
for both inflow and outflow at different parts of the star's surface.
In the models with magnetosphere, the star remaining within the mesh allows one
to study the effects of a spinning dipole. The disc surface is not
an outflow boundary condition, i.e.\ a region where mass is being injected
into the corona, but the outflow develops itself. Resolving the disc
makes it possible to include a disc dynamo which is the innovative ingredient
in the models of Paper~I and the present paper. Our modelling of the disc dynamo
will be explained in the following section.

We assume that the magnetic field in the disc be generated
by a standard $\alpha^2\Omega$ dynamo (e.g.,\ Krause \& R\"adler 1980),
where $\alpha$ is the mean-field $\alpha$ effect and
$\Omega$ is the angular velocity of the plasma.
This implies an extra electromotive force, $\alpha\BB$, in the induction
equation for the mean magnetic field, $\BB$, that is restricted to the disc.
As usual, the $\alpha$ effect is antisymmetric about the midplane
with (see Paper~I)
\begin{equation}
\alpha=\alpha_0\,{z\over
z_0}\,{\xi_{\rm disc}(\vec{r})\over1+v_{\rm A}^2/c_{\rm s}^2},     \label{alpha}
\end{equation}
where $v_{\rm A}$ is the Alfv\'en speed based on the total magnetic field,
$\xi_{\rm disc}$ is a profile specifying the shape of the disc
(see Sect.~\ref{Sec-cool-hot}),
and $\alpha_0$ is a
parameter that controls the intensity of dynamo action.
We choose the $\alpha$ effect to be negative in the upper half of the disc
(i.e.\ $\alpha_0<0$),
consistent with results from simulations of accretion disc turbulence driven by the
magneto-rotational instability (Brandenburg et al.\ 1995,
Ziegler \& R\"udiger 2000).
The resulting magnetic field symmetry is roughly dipolar.
This symmetry is seen both in three-dimensional simulations of
accretion disc dynamos driven by turbulence from
the magneto-rotational instability and in solutions of the
$\alpha \Omega$ dynamo problem with the $\alpha$ effect negative
in the upper disc half, provided that the accretion disc is embedded
in a conducting corona (e.g.,\ Brandenburg et al.\ 1990,
Brandenburg 1998,
von Rekowski et al.\ 2000, Bardou et al.\ 2001).
Further, we include $\alpha$ quenching which leads to
the disc dynamo saturating at a level close to equipartition between
magnetic and thermal energies.
The (constant) magnetic diffusivity $\eta$ is finite in the whole domain,
and enhanced in the disc by turbulence.
We do not take into account that the diffusivity in the star might also
be enhanced due to convection but assume the same value in the star as
in the corona.

To ensure that $\BB$ is solenoidal, we solve the induction equation
in terms of the vector potential $\AAA$, where $\BB=\vec{\nabla}\times\AAA$.

We impose
regularity conditions on the axis ($\varpi=0$) and outflow boundary
conditions on $\varpi=\varpi_{\max}$ and $z=\pm z_{\max}$.

\subsection{A cool disc in a hot corona: The initial state}
\label{Sec-cool-hot}

A simple way to implement a cool, dense disc embedded in a hot, rarefied corona
without modelling the detailed physics of coronal heating is to prescribe
the specific entropy, $s$, such that $s$ is smaller within the disc and
larger in the corona. For a perfect gas this implies $p=e^{s/c_v}\varrho^\gamma$
(in a dimensionless form), with the polytrope parameter $e^{s/c_v}$ being a
function of position.
In the model considered here,
we have an intermediate value for the specific entropy within the star.
We prescribe the polytrope parameter to be unity in the corona and
less than unity in the disc and in the star, so we put
\begin{equation}
e^{s/c_p}
= 1-(1{-}\beta_{\rm disc})\xi_{\rm disc}-(1{-}\beta_{\rm star})\xi_{\rm star},
\label{Kdef}
\end{equation}
where $\xi_{\rm disc}$ and $\xi_{\rm star}$ are (time-independent) profiles specifying
the shapes of the disc and the star.
The free parameters, $0<\beta_{\rm disc},\beta_{\rm star}<1$, control
the entropy contrasts between disc and corona and between star and corona,
respectively.

In the absence of a magnetosphere,
our initial state is a hydrostatic equilibrium with no poloidal velocity,
assuming an initially non-rotating hot corona that is supported
by the pressure gradient.
Since we model a disc that is cool, the disc is mainly centrifugally supported,
and as a result it is rotating at slightly sub-Keplerian speed.

The temperature ratio between disc and corona is roughly $\beta_{\rm disc}$.
Assuming pressure equilibrium between disc and corona,
and $p\propto\rho T$ for a perfect gas, the corresponding density ratio is then
$\beta_{\rm disc}^{-1}$.
Thus, the entropy contrast between disc and corona chosen here
($\beta_{\rm disc} = 0.005$), leads to
density and inverse temperature ratios of $200:1$ between disc and corona.

A rough estimate for the initial toroidal velocity, $u_{\varphi0}$,
in the midplane of the disc
follows from the hydrostatic equilibrium as
$u_{\varphi0}\approx\sqrt{1-\beta_{\rm disc}}\vec{\uu}_{\rm K}$.
For $\beta_{\rm disc}=0.005$, the toroidal velocity is within 0.25\%
of the Keplerian speed.

The initial hydrostatic solution is an unstable
equilibrium because of the vertical
shear between the disc, star and corona (Urpin \& Brandenburg 1998).
In addition, angular momentum transfer by viscous and magnetic
stresses -- the latter from the disc dynamo --
drives the solution immediately away from the initial state.

\subsection{Application to a protostellar star--disc system} \label{DVCP}

Since we use dimensionless variables, our model can be rescaled
and can therefore be applied to a range of different astrophysical objects.
Here, we consider values for our normalization parameters that are
typical of a protostellar star--disc system.
We scale the sound speed with a typical coronal sound speed of
$c_{\rm s0} = 10^2\km\s^{-1}$,
which corresponds to a temperature of $T_0 \approx 4\times10^5\K$,
and the disc surface density with $\Sigma_0=1\g\cm^{-2}$.
Further, we assume $M_*=1\,M_\odot$ (where $M_*$ and $M_\odot$ are the
stellar and solar mass, respectively), a mean specific weight of $\mu=0.6$,
and $\gamma=5/3$. This fixes the units of all quantities.
The resulting velocity unit is $[\vec{\uu}]=c_{\rm s0} = 10^2\km\s^{-1}$,
the length unit is $[\vec{\rr}]=GM_*/[\vec{\uu}]^{2}\approx0.1\AU$,
the time unit is $[t]=[\vec{\rr}]/[\vec{\uu}]\approx1.5~{\rm d}$,
the unit for the kinematic viscosity and magnetic diffusivity
is $[\nu] = [\eta] = [\vec{\uu}][\vec{\rr}] \approx 1.5 \times 10^{19} \cm^{2}\s^{-1}$,
the unit for specific entropy is $[s]=c_p=\gamma/(\gamma-1){\cal R}/\mu
\approx 3.5\times10^8\cm^2\s^{-2}\K^{-1}$,
the unit for specific enthalpy is $[h] = [\vec{\uu}]^{2} = 10^4\km^2\s^{-2}$,
the temperature unit is $[T] = [h]/[s] \approx 3\times10^5\K$,
the density unit is $[\varrho]=\Sigma_0/[\vec{\rr}]\approx7.5\times10^{-13}\g\cm^{-3}$,
the pressure unit is
$[p]=(\gamma-1)/\gamma[\varrho][h]\approx 30\g\cm^{-1}\s^{-2}$,
the unit for the mass accretion rate is
$[\dot{M}]=\Sigma_0 [\vec{\uu}] [\vec{\rr}] \approx 2\times10^{-7}M_\odot\yr^{-1}$,
the magnetic field unit is
$[\vec{\BB}]=[\vec{\uu}](4\pi[\varrho])^{1/2}\approx30\G$,
and the unit for the magnetic vector potential is
$[\vec{\AAA}]=[\vec{\BB}] [\vec{\rr}] \approx 4\times10^{13}\G\cm$.

Computations have been carried out in the domain
$(\varpi,z)\in[0,2]\times[-1,1]$, with the mesh sizes
$\delta \varpi=\delta z=0.01$. In our units, this corresponds to the domain
extending to $\pm 0.1\AU$ in $z$ and $0.2\AU$ in $\varpi$.

We choose a set of values for our model parameters such that
the resulting dimensions of our system and the resulting initial profiles
of the physical quantities
are close to those for a standard
accretion disc around a protostellar object. The stellar radius is $r_*=0.15$,
corresponding to $3$ solar radii, the disc inner radius is
$\varpi_{\rm in}=0.6$, i.e.\
$4$ stellar radii, the disc outer radius is $\varpi_{\rm out}=1.9$,
close to the outer domain
boundary, and the disc semi-thickness is $z_0=0.15$, i.e.\ equal to the stellar
radius.

Further, our choice for the entropy contrast between disc and corona,
$\beta_{\rm disc} = 0.005$, leads to an initial disc temperature
ranging between $9000\K$ in the inner part and $900\K$ in the outer part.
Real protostellar discs have typical temperatures
of about a few thousand Kelvin (e.g.,\ Papaloizou \& Terquem 1999).
As turns out from our simulations, the disc temperature increases by less than
a factor of $2$ with time, except for the inner disc edge where the increase
is much higher.
The low disc temperature corresponds to a relatively high disc density.
In the inner part of the disc, it is about $3\times10^{-10}\g\cm^{-3}$
initially, but increases to about $10^{-9}\g\cm^{-3}$ at later stages.
In the outer part it is a few times $10^{-11}\g\cm^{-3}$.

The Shakura--Sunyaev coefficient of the turbulent disc viscosity is
$\alpha_{\rm SS} = 0.004$ whereas the magnetic diffusivity is $2\times 10^{-5}$
in the corona and in the star, and enhanced to $6\times 10^{-5}$
in the disc. The latter corresponds to
$\alpha_{\rm SS}^{(\eta)} \equiv {\eta_{\rm disc}/(c_{\rm s,disc}z_0})$,
ranging between $0.004$ and $0.008$ in the disc midplane.
With $\eta_{\rm disc} = 6\times 10^{-5}$ (corresponding to about
$9 \times 10^{14} \cm^{2}\s^{-1}$) and the disc semi-thickness $z_0=0.15$,
the diffusion time is $t_{\rm diff} \equiv z_0^2/\eta_{\rm disc} = 375$.

The stellar surface angular velocity is about unity
(in dimensionless units), corresponding to a rotation period of around 10~days.
This means that the corotation radius is $\varpi_{\rm co}\approx 1$
in nondimensional units.
The inner disc radius (which is fixed in our model) is rotating with
roughly Keplerian speed, resulting in a rotation period of the inner disc edge
of around 4.5~days (3 in dimensionless units).

\subsection{Disc dynamo and stellar magnetosphere} \label{ADDSM}

To date, almost all models of the formation and collimation
of winds and jets from protostellar accretion discs rely on an externally
imposed poloidal magnetic field and ignore any field produced in the disc.
This is not the case in the model developed in Paper~I, which also
forms the basis of the model used in the present paper. In Paper~I we
study outflows
in connection with magnetic fields that are solely generated and
maintained by a disc dynamo,
resolving the disc and the star at the same time as well as assuming a non-ideal
corona.
There, the initial seed magnetic field is large scale, poloidal,
of mixed parity, weak and confined to the disc.

However, observations of T~Tauri stars suggest that magnetic star--disc coupling
might result in a spin-down of the star due to magnetic braking by a stellar
magnetosphere penetrating the disc. Magnetic fields as strong as $1\kG$ and
larger have been detected on T~Tauri stars (e.g.,\ Guenther et al.\ 1999),
and there is evidence for hot and cool spots on the stellar surface.

Our present model is similar to that of Paper~I except that,
in addition to the disc dynamo, we also
model a magnetosphere of the protostar. For the magnetosphere, we assume that
in the initial state a stellar dipolar magnetic field threads
the surrounding disc, leading to a contribution of $A_\varphi$ given by
\begin{equation}
A_\varphi(\varpi,z)= \dots +
A_{\rm star} {\varpi r_*^2 \over r^3} \left(1-\xi_{\rm star}\right),
\label{Eqmagsphere}
\end{equation}
where $r_*$ is the stellar radius,
$r=(\varpi^2+z^2)^{1/2}$ is the spherical radius, and $A_{\rm star}$ is a
parameter controlling the strength of the stellar magnetosphere.
Since $\varpi=r\sin\Theta$ (with $\Theta$ the co-latitude), the magnetic moment
is $\mu_{\rm mag} = A_{\rm star} r_*^2$ (cf.\ Model~I in Miller \& Stone 1997).
The dots in Eq.~(\ref{Eqmagsphere}) denote the contribution
from the seed magnetic field in the disc.
Since the magnetic field describing the magnetosphere is force-free,
the hydrostatic equilibrium is not affected but equal to the magnetostatic
equilibrium.
$\xi_{\rm star}$ is a smoothed profile for the star and equal to unity only
at the origin (centre of the star), so that the magnetosphere extends
to parts of the star as well.
In order to anchor the magnetosphere within the star and on the stellar surface,
$\varrho$, $\uu$, and $\AAA$ are set to their initial values at each timestep
in the region including the star and extending to about $1.5 r_*$.
Therefore, in the models with magnetosphere, there is no mass sink
in the star of the type discussed in Paper~I. The star is resolved in that it stays
within the mesh, allowing the study of the effects of a spinning dipole
(cf.\ Sect.~\ref{BE}).
Note that outside the anchoring region, the magnetosphere
is not imposed and can evolve dynamically with time. In the induction equation,
we do not include any term containing an external magnetic field.
As a reference model we choose $A_{\rm star} = 5$,
which corresponds to a stellar surface magnetic field strength of about $1\kG$.
The magnetic moment is then about $10^{37}\G\cm^3$.

\section{Results} \label{RES}

We consider models with magnetospheres of different strengths.
We declare as our reference model one with a stellar surface field
strength of $1\kG$ (Sect.~\ref{RM}), and compare it with a model
without disc dynamo (Sect.~\ref{NDD}), and with models whose
stellar surface field strength is varied between zero and $5\kG$
(Sects~\ref{WM}--\ref{nomag}).
Finally, the magnetic and accretion torques on the star are studied
for different magnetospheric models (Sect.~\ref{MAAT}).

\subsection{Reference model:~ $|B_{\rm surf}| \approx 1\kG$ (Model~M1)}
\label{RM}

\begin{figure}[t!]
\centerline{
   \includegraphics[width=8.5cm]{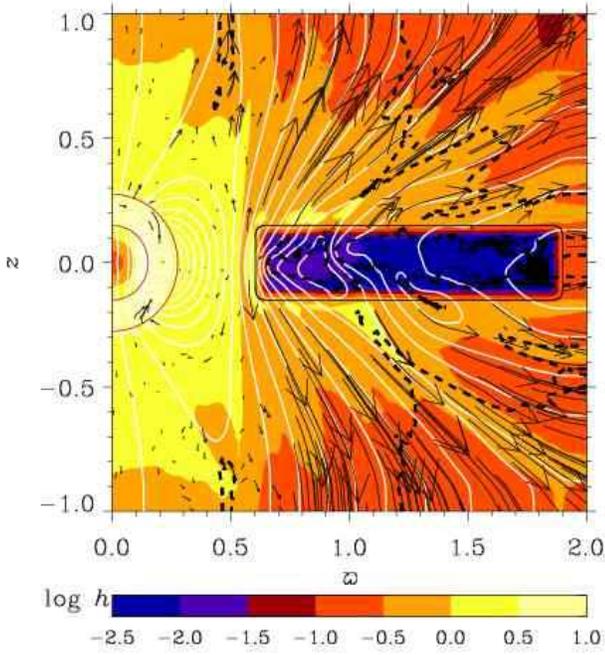}
}
\caption[]{
Poloidal velocity vectors and poloidal magnetic field lines (white)
superimposed on a colour/grey scale representation of $\log_{10}h$ for
our reference model (Model~M1)
with a stellar surface magnetic field strength of
about $1\kG$
and with disc dynamo at the time $t=1267$.
Specific enthalpy
$h$ is directly proportional to temperature $T$, and $\log_{10}h=(-2,-1,0,1)$
corresponds to $T\approx (3{\times}10^{3},3{\times}10^{4},3{\times}10^{5},
3{\times}10^{6})\,\mbox{K}$. The black dashed line shows the surface where the
poloidal velocity equals the Alfv\'en speed from the poloidal magnetic field
(Alfv\'en surface).
The disc boundary is shown as a thin black line. In red are marked the stellar
surface as well as the surface up to which the magnetosphere is anchored.
Here and in the following figures, the field lines tend to become
vertical near the boundaries, but this is due to artifacts from the boundary
conditions.
}\label{FRun_mag_medium1-pnew-1267}
\end{figure}

Our simulations show that,
similar to Paper~I, also the present model with a stellar magnetosphere
(in addition to the disc dynamo)
develops a structured outflow, composed of a stellar wind
and a disc wind; see Fig.~\protect\ref{FRun_mag_medium1-pnew-1267}.
The disc wind is faster, cooler and less dense, with the highest velocities
in the wind originating from the inner edge of the disc,
whereas the stellar wind is slower, hotter and denser.
The inner disc wind is magneto-centrifugally accelerated,
whereas the outer disc wind and the stellar wind are mostly pressure-driven.
These driving and acceleration mechanisms will be explained in Sect.~\ref{SM},
using Model~S in which the outflow structure is most pronounced.

The disc wind velocity reaches about $240\km\s^{-1}$,
whereas the stellar wind velocity goes only up to about $10\km\s^{-1}$.
The disc wind mass loss rate is time-dependent with an average value of
around $2\times 10^{-7}M_\odot\yr^{-1}$.

\begin{figure}[t!]
\centerline{
   \includegraphics[width=8.5cm]{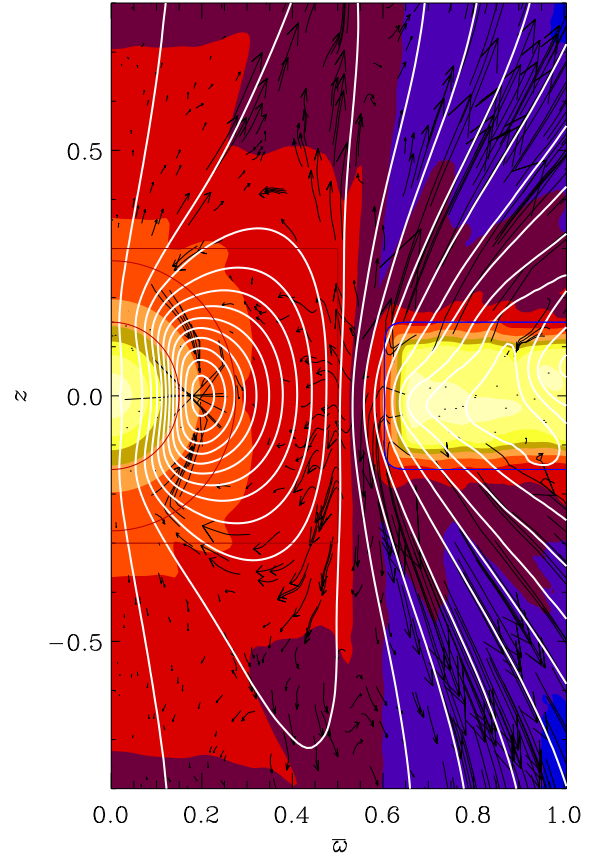}
}
\caption[]{
Reference model (Model~M1) at the time $t=1267$ when the accretion rate is maximum.
Colour/grey scale representation of the density (bright colours or light shades
indicate high values; dark colours or dark shades indicate low values)
with poloidal magnetic field lines superimposed (white) and
the azimuthally integrated mass flux density, represented as the vector
$2\pi\varpi\varrho(u_\varpi,u_z)$, shown with arrows
(except in the disc where the density is high and the mass flux vectors
would be too long).
}\label{FRun_mag_medium1-p-1267}
\end{figure}

Accretion of matter onto the central star is highly episodic with a maximum rate of
about $4\times 10^{-9}M_\odot\yr^{-1}$. Figure~\protect\ref{FRun_mag_medium1-p-1267}
shows a snapshot at a time when the accretion rate is maximum.
We estimate the mass accretion rate in the following way.
We put a cylinder around the axis extending to $\varpi=0.5$ and $z=\pm 0.3$.
(Note that the disc inner edge is at $\varpi=0.6$, and the magnetosphere
is anchored up to $r_0 \approx 0.275$ due to a smoothed profile for the
anchoring region.)
Then we calculate the azimuthally integrated mass flux density,
$2\pi\varpi\varrho\uu$,
normal to the boundaries of the cylinder and integrate it over the boundaries.
We take into account that there is mass loss due to the stellar wind.
It turns out that in this model, matter enters the cylinder mainly through
the vertical boundary and only then partly flows along magnetospheric field lines,
because the magnetosphere does not extend far enough.

\subsection{A model with no disc dynamo (Model~M1-0)} \label{NDD}

\begin{figure}[t!]
\centerline{
   \includegraphics[width=8.5cm]{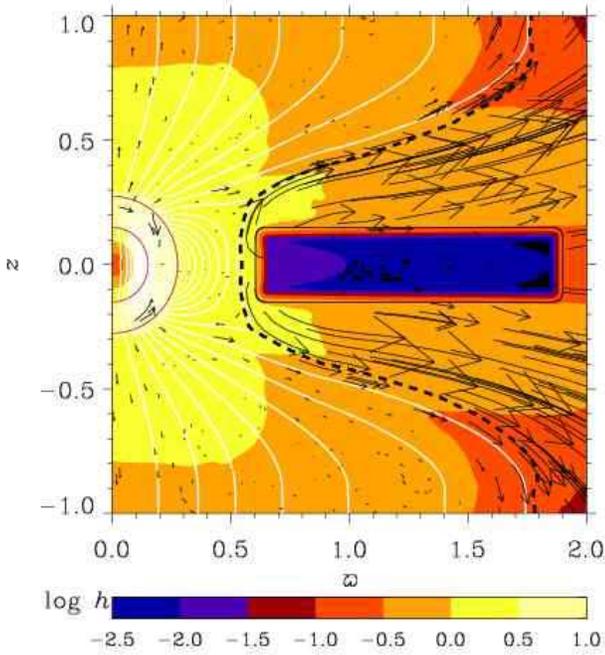}
}
\caption[]{
Model~M1-0.
Same model as the reference model in Fig.~\protect\ref{FRun_mag_medium1-pnew-1267},
but with no disc dynamo, i.e.\ $\alpha=0$ in \Eq{Ind}, and
averaged over times $t=1100 \dots t=1127$.
Note that, unlike all the models with disc dynamo, the magnetic field
(field lines shown in white) is almost entirely swept out of the disc.
}\label{FRun_mag_medium1-nodyn-pnew-AVER800_827}
\end{figure}

Almost all previous work on the magnetic star--disc
coupling assumes that the magnetic field in the disc results entirely
from the central star. Some models include an externally imposed
magnetic field (e.g.,\ K\"uker et al.\ 2003).
In order to study the effect of the disc dynamo on the overall field
structure and on the resulting outflow, we have calculated a model
where the disc dynamo is turned off,
i.e.\ we put $\alpha=0$ in \Eq{Ind}.

As can be seen in Fig.~\protect\ref{FRun_mag_medium1-nodyn-pnew-AVER800_827},
the magnetic field in the disc is now almost entirely swept away. The initial
disc field due to the penetrating magnetosphere is expelled
and only a very weak disc field remains without a dynamo.
As a consequence, the structure and driving/acceleration mechanism
of the disc wind change significantly.
Transport of specific angular momentum is weak and almost entirely
in the $\varpi$ direction,
along very weak field lines threading the disc and running in the low corona
almost parallel to the disc surface. The resulting disc wind along these
field lines is therefore also mainly in the horizontal direction
(Fig.~\protect\ref{FRun_mag_medium1-nodyn-pnew-AVER800_827}).
Since the field threading the disc is very weak, the disc wind is always
super-Alfv\'enic, reaching a speed of about $160\km\s^{-1}$,
so that no magneto-centrifugal acceleration can take place.
Characteristic for the models with disc dynamo is a
cooler, less dense region (``conical shell'') that originates
from the inner disc edge, and that has higher specific angular momentum
than elsewhere
and contains the inner magneto-centrifugally accelerated disc wind.
This conical shell disappears as the magnetic field, that was threading
the disc, is swept away and no disc dynamo is generating and maintaining
a strong enough field. As a result, the pressure force becomes then the only
driving mechanism of the whole disc wind.

As the disc field is expelled, the size of the closed stellar magnetosphere shrinks;
the extended open stellar field lines now find place to run in the corona
parallel above the very weak field lines threading the disc;
see Fig.~\protect\ref{FRun_mag_medium1-nodyn-pnew-AVER800_827}.
[Note that in our magnetospheric models, the vertical alignment of the field
lines near the boundaries
is due to boundary conditions. In order to stabilize our code at the boundaries
(especially in situations where the wind speed is low),
we force matter to be slowly advected from the computational domain normal
to the boundaries, rather than in spherical radial direction as in Paper~I.]
The stellar wind, however, does not seem to be
much affected, with a maximum speed of about $10\km\s^{-1}$.

The average disc wind mass loss rate does not change much in comparison
with the reference model, and remains
around $2\times 10^{-7}M_\odot\yr^{-1}$.
Accretion is still highly episodic; the maximum mass accretion rate
is roughly 5 times larger
compared to the same model but with disc dynamo, so it is now
about $2\times 10^{-8}M_\odot\yr^{-1}$.

\subsection{Weak magnetosphere:~ $|B_{\rm surf}| \approx 200\G$ (Model~W)}
\label{WM}

\begin{figure}[t!]
\centerline{
   \includegraphics[width=8.5cm]{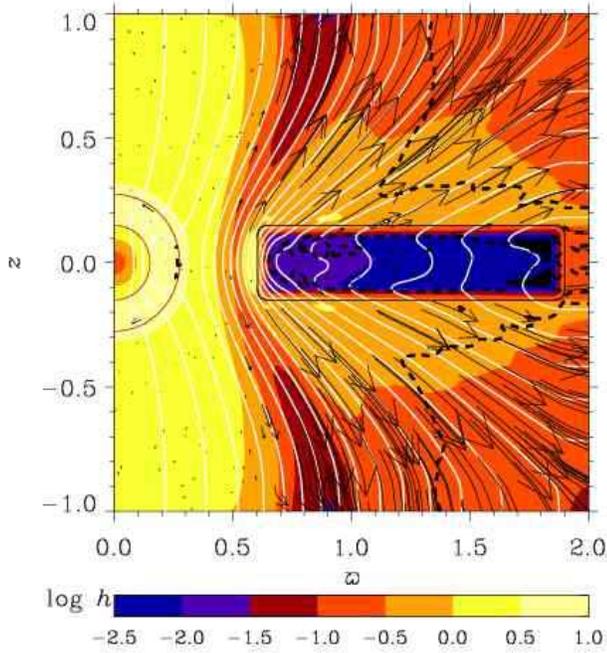}
}
\caption[]{
Model~W (weak magnetosphere and disc dynamo); the magnetic field strength
at the stellar surface is about $200\G$.
Averaged over times $t=825 \dots t=850$.
Shown is the same as in Fig.~\protect\ref{FRun_mag_medium1-pnew-1267}.
}\label{FRun_mag_weak-pnew-AVER1785_1835}
\end{figure}

A stellar surface field strength of $200\G$ is not large enough to maintain a closed
magnetosphere; see Fig.~\protect\ref{FRun_mag_weak-pnew-AVER1785_1835}.
The field lines open up entirely to form open stellar and disc fields.
This mainly affects the stellar wind; its maximum speed is reduced to about
$2.5\km\s^{-1}$, whereas the disc wind velocity reaches about $200\km\s^{-1}$.
However, the disc dynamo again produces a structure in the disc wind
with different driving mechanisms.
Transport of specific angular momentum is also enhanced along field lines
threading
the midplane at a radius smaller than the prescribed inner disc radius,
which is due to an increased angular velocity.
In the disc, the field lines show a characteristic wiggly pattern similar
to the so-called channel flow solution seen in two-dimensional simulations
of the Balbus-Hawley instability (Hawley \& Balbus 1991).
This instability is indeed to be expected for the weak field strengths
in the present model; for stronger fields this instability
is suppressed.
Note, however, that this channel flow type solution in the disc can only
be seen in the {\it time-averaged} picture shown in
Fig.~\protect\ref{FRun_mag_weak-pnew-AVER1785_1835}.

\begin{figure}[t!]
\centerline{
   \includegraphics[width=8.5cm]{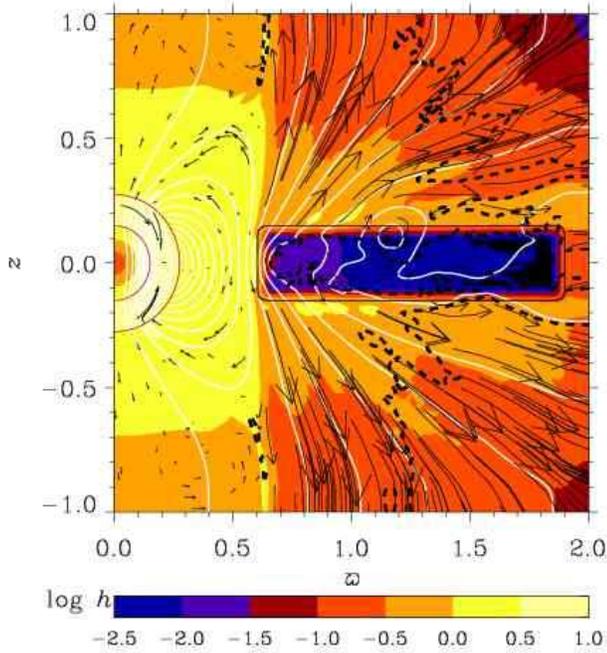}
}
\caption[]{
Model~M2 (medium magnetosphere and disc dynamo); the magnetic field strength
at the stellar surface is about $2\kG$.
$t=562$.
Shown is the same as in Fig.~\protect\ref{FRun_mag_medium1-pnew-1267}.
}\label{FRun_mag_medium2-pnew-562}
\end{figure}

\begin{figure}[t!]
\centerline{
   \includegraphics[width=8.5cm]{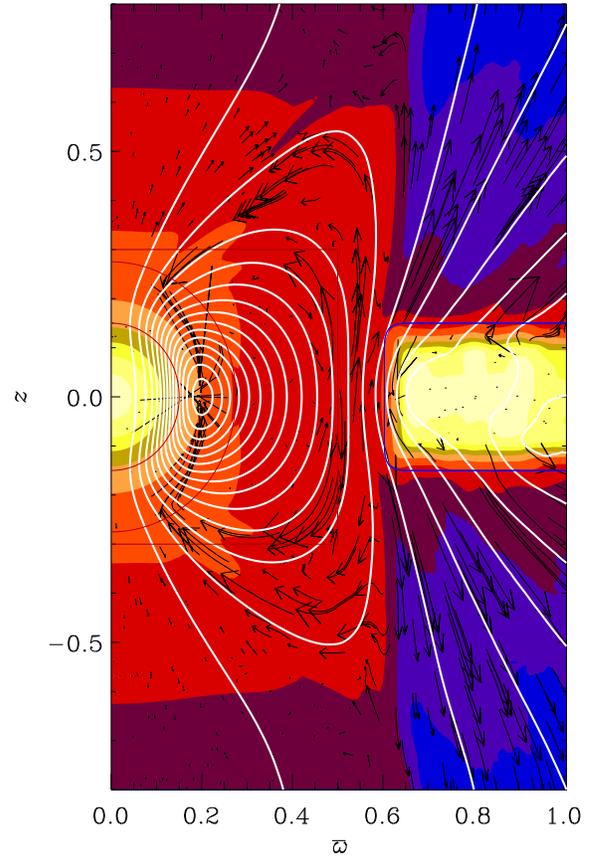}
}
\caption[]{
Model~M2.
$t=562$.
Shown is the same as in Fig.~\protect\ref{FRun_mag_medium1-p-1267},
also at a time when the accretion rate is maximum.
}\label{FRun_mag_medium2-p-562}
\end{figure}

In this connection it might be worthwhile reiterating that the purpose
of introducing a turbulent viscosity is of course to imitate the effects
of the three-dimensional Balbus-Hawley instability.
The channel flow solution is clearly an artifact of two dimensions, and
we regard the almost complete absence of the channel flow behaviour in our
solutions as a confirmation that there is no ``double-instability'' of the
mean flow in the sense discussed above [see also Tuominen et al.\ (1994) and
Yousef et al.\ (2003) for discussions regarding the problems of stability
studies of mean-field models].

\subsection{Medium magnetosphere:~ $|B_{\rm surf}| \approx 2\kG$ (Model~M2)}
\label{MM}

A stellar magnetosphere with an increased stellar surface field strength
of about $2\kG$
interacts with the disc dynamo in such a way that the resulting
magnetic field, temperature and density distributions are clearly influenced.
The structure in the disc wind becomes more visible, with a more pronounced
(cooler and less dense) conical shell originating from the inner disc edge
containing a (slightly) faster magneto-centrifugally accelerated
inner disc wind with a maximum speed of about $250\km\s^{-1}$; compare
Figs~\protect\ref{FRun_mag_medium1-pnew-1267} and \ref{FRun_mag_medium2-pnew-562}.
The average disc wind mass loss rate stays around $2\times 10^{-7}M_\odot\yr^{-1}$,
and the stellar wind velocity does not exceed $10\km\s^{-1}$.

When the stellar surface field is as strong as $2\kG$,
the highly episodic accretion is correlated with magnetospheric oscillations.
During time intervals when the magnetosphere is expanded so that the
outer closed field lines are sufficiently close to the disc,
the accretion flow is along magnetospheric field lines;
see Fig.~\protect\ref{FRun_mag_medium2-p-562} (see also Sect.~\ref{SM}).
The maximum accretion rate is about $2\times 10^{-8}M_\odot\yr^{-1}$.

\subsection{Strong magnetosphere:~ $|B_{\rm surf}| \approx 5\kG$ (Model~S)}
\label{SM}

\begin{figure}[t!]
\centerline{
   \includegraphics[width=8.5cm]{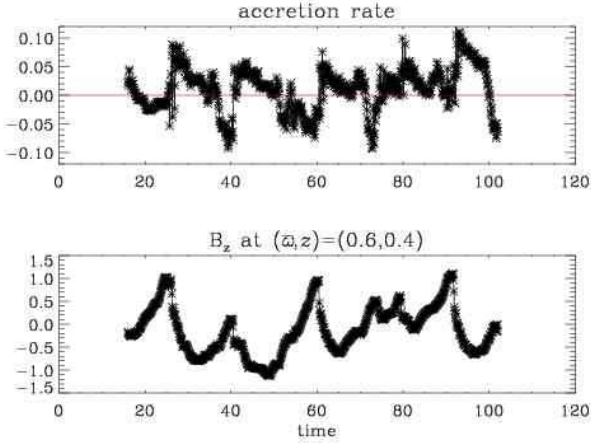}
}
\caption[]{
Model~S (strong magnetosphere and disc dynamo); the magnetic field strength
at the stellar surface is about $5\kG$.
Mass accretion rate estimated as described in Sect.~\ref{RM}
(upper panel) and vertical magnetic field component at $(\varpi,z)=(0.6,0.4)$
(lower panel) as functions of time. The latter is an indicator for the oscillations
of the magnetosphere. Note that the accretion rate and the magnetospheric
oscillations are correlated such that the accretion rate is high when
$B_z$ at the position given above is low.
This corresponds to the state when the star is connected to the disc by
its magnetosphere.
Time is here in non-dimensional units, but with a time unit of 1.5~days,
the oscillation period is seen to be around 15 to 30~days.
}\label{pfaveracc_star-two_Bz}
\end{figure}

\begin{figure}[t!]
\centerline{
   \includegraphics[width=8.5cm]{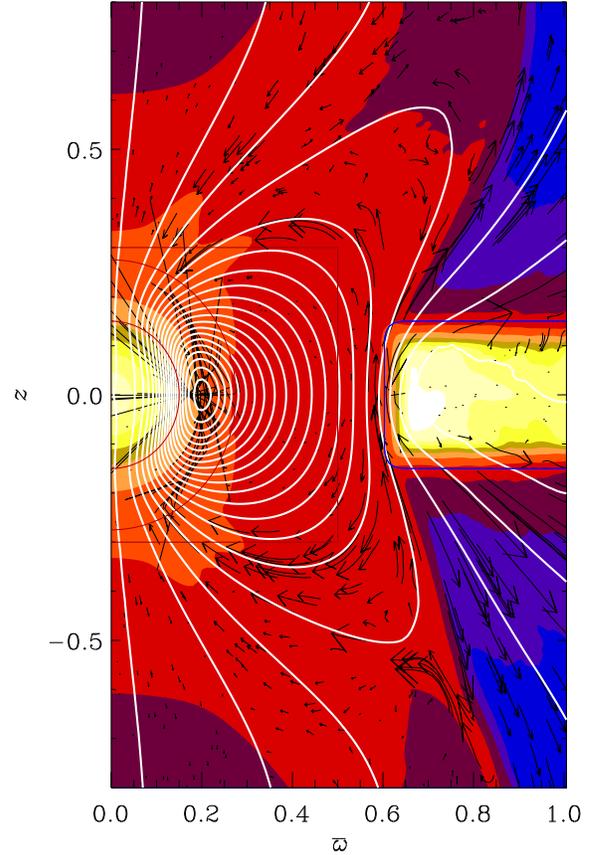}
}
\caption[]{
Model~S.
Shown is the same as in Fig.~\protect\ref{FRun_mag_medium1-p-1267},
but at the time $t=98$ when the star is magnetically connected
to the disc.
}\label{FRun_mag_strong-p-98}
\end{figure}

\begin{figure}[t!]
\centerline{
   \includegraphics[width=8.5cm]{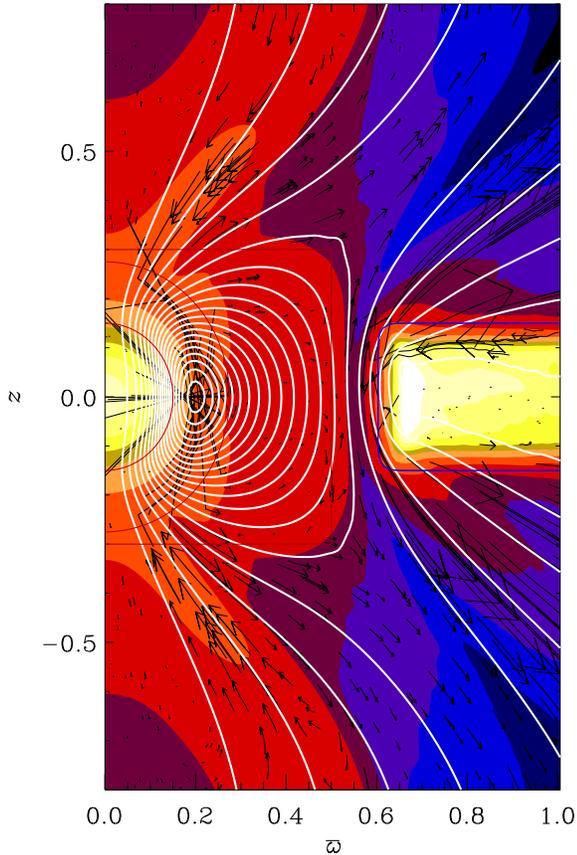}
}
\caption[]{
Model~S.
Shown is the same as in Fig.~\protect\ref{FRun_mag_strong-p-98},
but at the time $t=102$ when the star is magnetically disconnected
from the disc.
}\label{FRun_mag_strong-p-102}
\end{figure}

The time sequence shown in the lower panel of
Fig.~\protect\ref{pfaveracc_star-two_Bz} shows that
the magnetosphere is oscillating with a period of around 15 to 30~days.
As we shall show below,
the oscillations mean that the magnetosphere is expanding and contracting
and in this way periodically connecting and disconnecting open stellar
and disc fields, therefore changing the magnetic star--disc coupling.

It turns out that the highly episodic accretion flow and accretion rate are
correlated with the configuration of the magnetosphere
(compare upper and lower panels of Fig.~\protect\ref{pfaveracc_star-two_Bz}).
Low values of $B_z$ at the position $(\varpi,z)=(0.6,0.4)$, correspond to
a configuration where closed magnetospheric lines penetrate the inner disc edge,
thus connecting the star to the disc
(as in Fig.~\protect\ref{FRun_mag_strong-p-98}).
In this case, disc matter is loaded onto magnetospheric field lines
and flows along them to accrete onto the star.
As a consequence, the accretion rate is highest during these phases,
typically up to between
$10^{-8}M_\odot\yr^{-1}$ and $2.5\times10^{-8}M_\odot\yr^{-1}$
(see Fig.~\protect\ref{pfaveracc_star-two_Bz}, upper panel).

High values of $B_z$ at the position $(\varpi,z)=(0.6,0.4)$, correspond to
a configuration when the outer field lines of the magnetosphere have opened up
into disconnected open stellar and disc field lines,
thus disconnecting the star from the disc
(as in Fig.~\protect\ref{FRun_mag_strong-p-102}).
In this case, matter is lost directly into the outflow and
there is no net accretion of disc matter
(see Fig.~\protect\ref{pfaveracc_star-two_Bz}, upper panel).

\begin{figure}[t!]
\centerline{
   \includegraphics[width=8.5cm]{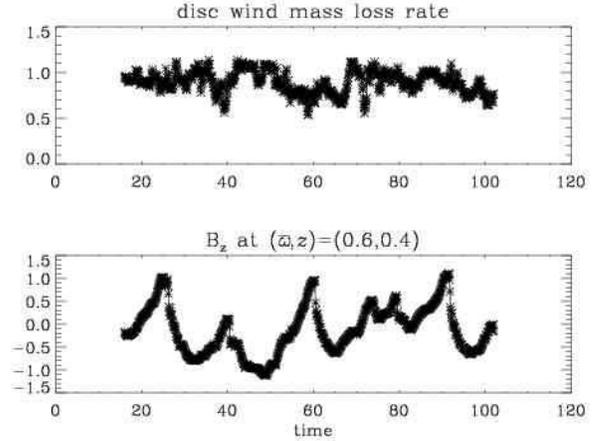}
}
\caption[]{
Model~S.
Disc wind mass loss rate (upper panel) and
vertical magnetic field component at $(\varpi,z)=(0.6,0.4)$ (lower panel)
as functions of time. The latter is an indicator for the oscillations
of the magnetosphere. No correlation between the disc wind mass loss rate
and the magnetospheric oscillations can be seen.
}\label{pfaveracc_disc-two_Bz}
\end{figure}

However, even when the accretion rate is high, most of the mass that leaves the
disc goes into the wind. The mass loss rates into the disc wind are about one
order of magnitude higher compared to the peak accretion rates
(see Fig.~\protect\ref{pfaveracc_disc-two_Bz}, upper panel), and not very
different from the rates in our other magnetospheric models
presented in previous subsections,
ranging between $10^{-7}M_\odot\yr^{-1}$ and
$2.5\times10^{-7}M_\odot\yr^{-1}$.
The disc wind mass loss rates are fluctuating with time on rather short time
scales, and no correlation with the oscillating magnetosphere can be detected
(compare upper and lower panels of Fig.~\protect\ref{pfaveracc_disc-two_Bz}).

The time-dependent behaviour of the accretion flow and the disc wind is
interesting in view of observations. Recent high-resolution short- and long-term
monitoring of
classical T~Tauri stars (CTTS) with the ESO Very Large Telescope reveals that
CTTS are highly variable on time scales of minutes to several years. This
variability in the emission spectra is associated with both accretion and
outflow processes (Stempels \& Piskunov 2002, 2003).

\begin{figure}[t!]
\centerline{
   \includegraphics[width=8.5cm]{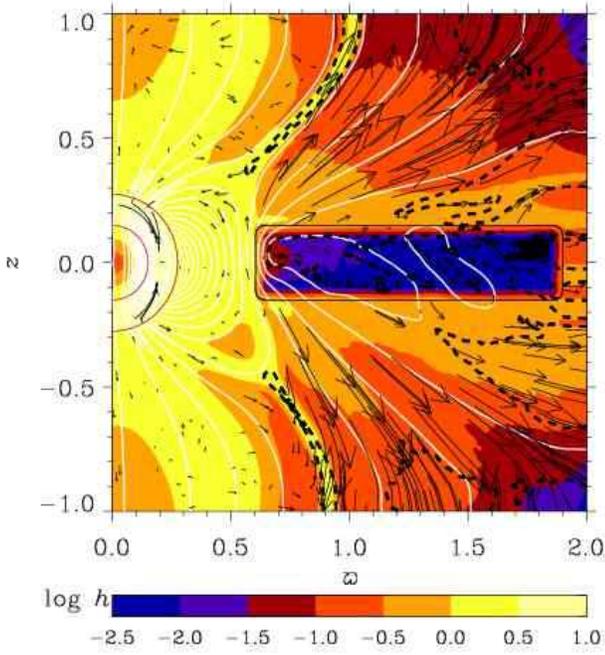}
}
\caption[]{
Model~S.
$t=103$ when the star is about to be magnetically reconnected to the disc.
Shown is the same as in Fig.~\protect\ref{FRun_mag_medium1-pnew-1267}.
}\label{FRun_mag_strong-pnew-103}
\end{figure}

As in all our models with disc dynamo, the dynamo produces a structured disc wind.
However, the larger the stellar surface magnetospheric field strength is,
the more pronounced is the structure.
The outflow consists of (i) a slower, hotter and denser, mostly
pressure-driven stellar wind,
(ii) a faster, cooler and less dense outer disc wind that is also mostly
pressure-driven, and (iii) a fast magneto-centrifugally accelerated
inner disc wind within a clearly visible conical shell originating from the
inner disc edge, that has lower temperatures and densities than elsewhere (see
Fig.~\protect\ref{FRun_mag_strong-pnew-103}).
This outflow structure was already discussed in Paper~I in some detail.

Figure~\protect\ref{FRun_mag_strong-pnew-103} shows a snapshot at a time
in a transition period, when the magnetic star--disc connectivity changes.
In these periods when disconnected stellar and disc field lines are about
to reconnect, in addition to the above described structured outflow
typical for our models with disc dynamo,
there is a pressure-driven, hot and dense, but relatively fast
outflow between the stellar and disc winds.

\begin{figure}[t!]
\centerline{
   \includegraphics[width=8.5cm]{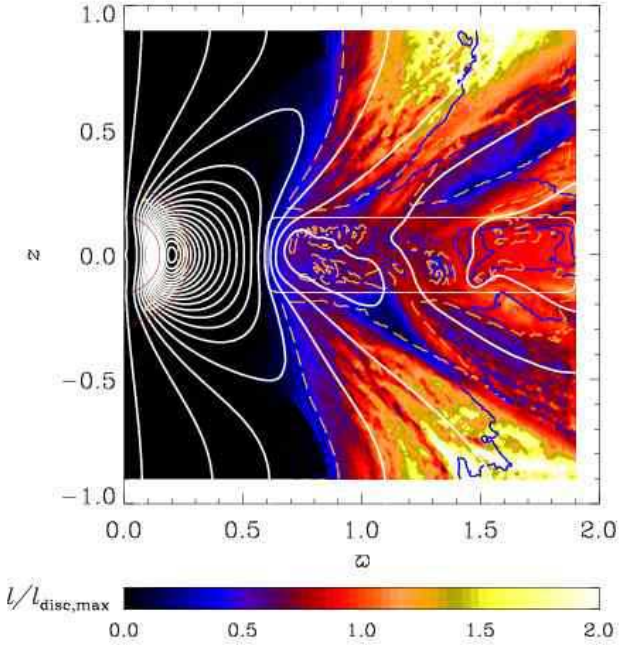}
}
\caption[]{
Colour/grey scale representation of the specific angular momentum,
$l = \varpi u_\varphi = \varpi^2 \Omega$,
normalized by the maximum specific angular momentum in the disc,
$l_{\rm disc,max}$,
with poloidal magnetic field lines superimposed (white)
for Model~S at the time
$t=98$ when the star is magnetically connected to the disc.
The blue solid line shows the Alfv\'en surface, and the orange dashed line
the sonic surface.
}\label{FRun_mag_strong-pangmom-98}
\end{figure}

\begin{figure}[t!]
\centerline{
   \includegraphics[width=8.5cm]{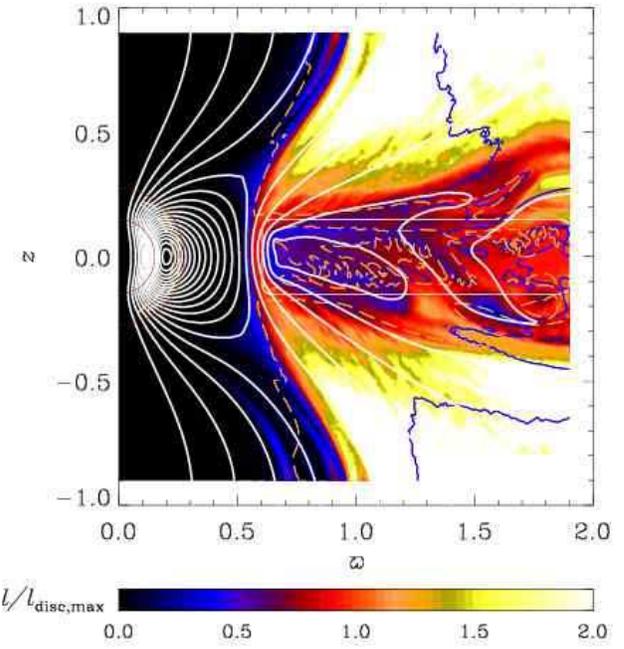}
}
\caption[]{
Same as in Fig.~\protect\ref{FRun_mag_strong-pangmom-98} (Model~S), but at
the time $t=102$ when the star is magnetically disconnected from the disc.
Note that now a lot of specific angular momentum leaves the disc from
its inner edge.
}\label{FRun_mag_strong-pangmom-102}
\end{figure}

As one can see in Figs~\protect\ref{FRun_mag_strong-pangmom-98} and
\ref{FRun_mag_strong-pangmom-102},
specific angular momentum is carried outwards from the disc into the corona
mainly
along magnetic field lines threading the innermost part of the disc.
Here the dynamo-generated field is strongest
and is responsible for confining the conical shell in the corona.
The angle between the rotation axis and these field lines
is $30^\circ$ and larger at the disc surface, which is favourable for
magneto-centrifugal acceleration (Blandford \& Payne 1982;
see Campbell 1999, 2000, 2001 for a more detailed treatment).
Furthermore, in the conical shell the flow is highly supersonic but
the Alfv\'en radius is about two times larger than
the radius where the field lines have their footpoints at the
disc surface.
This is sufficient for the magnetic field lines to act as a lever arm
along which the inner disc wind is accelerated.
On the other hand, in the outer disc wind the Alfv\'en surface is
very close to the
disc surface. As a consequence, acceleration due to the pressure gradient
becomes more important there, although the criterion of Blandford \& Payne
(1982) is still fulfilled.
The stellar wind is mostly pressure-driven, carrying almost no specific
angular momentum
and remaining subsonic.
The outflow between the stellar and disc winds has small azimuthal velocity
and small specific angular momentum, and becomes quickly super-Alfv\'enic.

\begin{figure}[t!]
\centerline{
   \includegraphics[width=8.5cm]{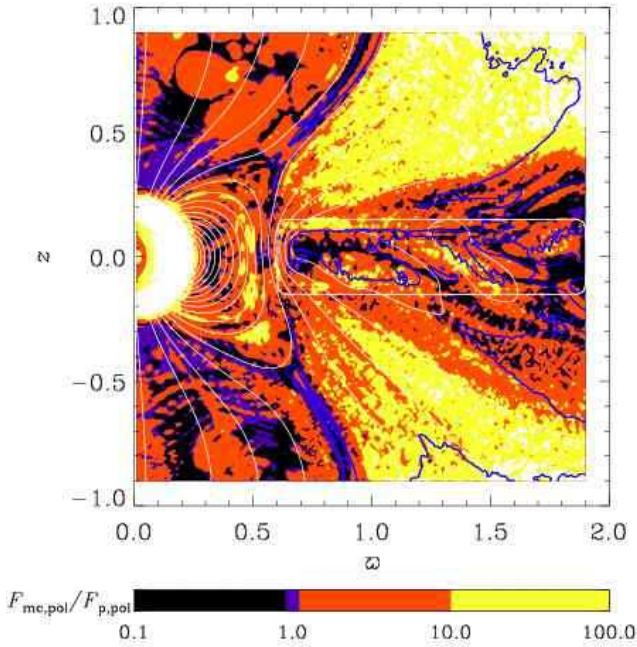}
}
\caption[]{
Colour/grey scale representation of the ratio
between the poloidal magneto-centrifugal and pressure forces,
$|\FF_{\rm pol}^{\rm(mc)}|/|\FF_{\rm pol}^{\rm(p)}|$ as defined in
Eq.~(\ref{PolForces}) and below, with larger values corresponding to
lighter shades.
Superimposed are the poloidal magnetic field lines (white). The blue solid line
shows the Alfv\'en surface.
Model~S at the time
$t=103$ when the star is about to be magnetically reconnected to the disc.
}\label{FRun_mag_strong-pforces-103}
\end{figure}

A confirmation of the presence of coexisting pressure-driving and
magneto-centrifugal acceleration of the winds as described above,
can be obtained by looking at the ratio
between the poloidal magneto-centrifugal and pressure forces,
$|\FF_{\rm pol}^{\rm(mc)}|/|\FF_{\rm pol}^{\rm(p)}|$, where
the subscript `pol' denotes the poloidal components.
The forces appear in the equation of motion as
\EQ
\FF^{\rm(mc)}=\varrho\left(\Omega^2\vec{\varpi}-\vec{\nabla}\Phi\right)
+\JJ\times\BB,
\label{PolForces}
\EN
and $\FF^{\rm(p)}=-\vec{\nabla}p$. Their ratio is shown in
Fig.~\protect\ref{FRun_mag_strong-pforces-103}. Again, the conical shell can be
clearly seen. In the conical shell, the ratio assumes large values which
confirms that magneto-centrifugal launching and acceleration of the disc wind
is dominant there.
On the other hand, at the outer parts of the disc surface the pressure force
is stronger, leading to pressure-driving. At the stellar surface, both forces
have comparable strengths; however, in large parts of the stellar wind
the pressure force becomes dominant.
The outflow between the stellar and disc winds is also pressure-dominated.

\begin{figure}[t!]
\centerline{
   \includegraphics[width=8.5cm]{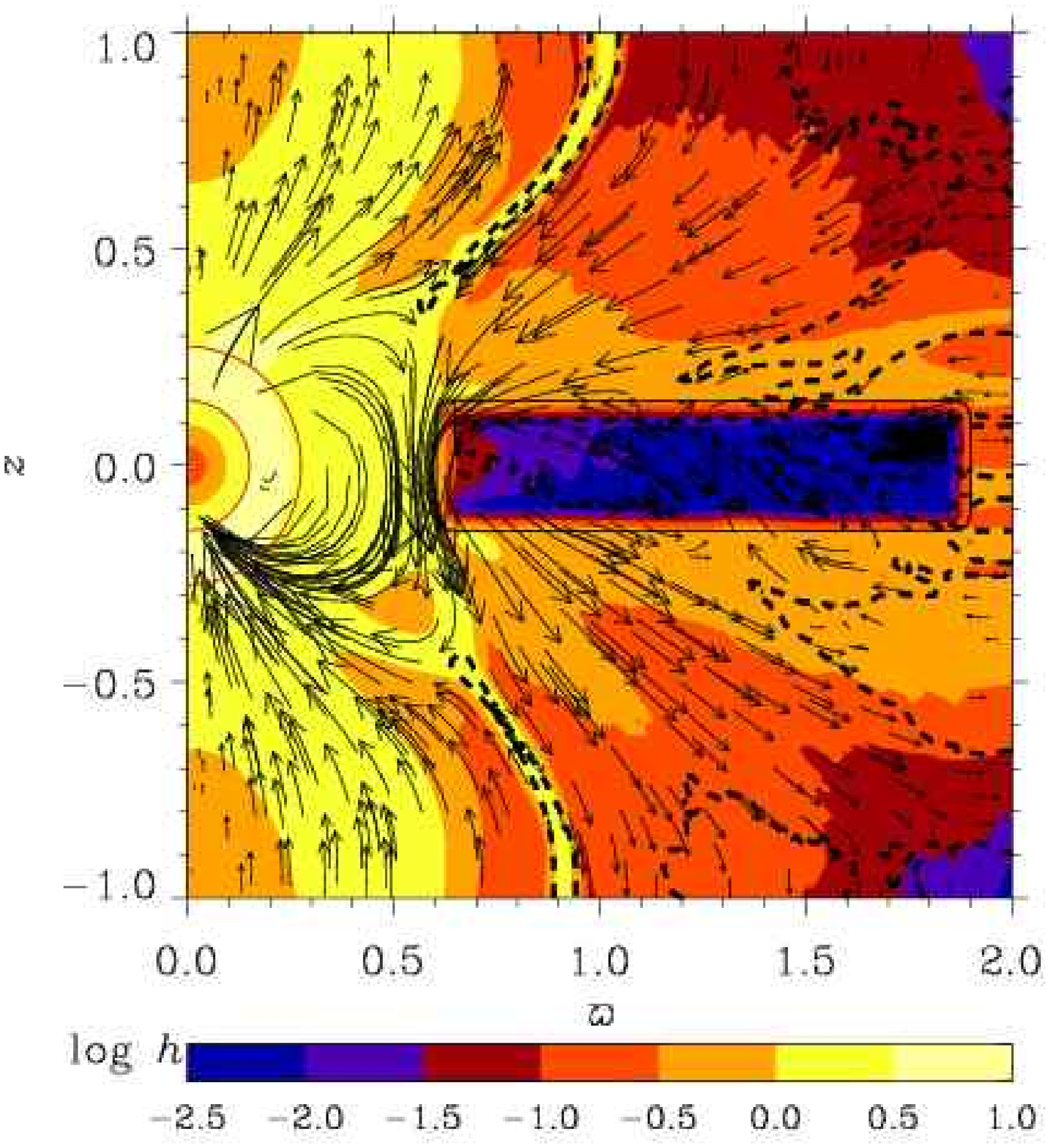}
}
\caption[]{
Poloidal magnetic field vectors of the model shown in
Fig.~\protect\ref{FRun_mag_strong-pnew-103}.
Note that the magnetic field reversal in the corona indicates the presence
of a current sheet between the magnetosphere and the field threading
the disc at about $45^\circ$. This means that no {\sf X}-point forms.
The magnetic field vectors are not shown in the anchoring region.
The length of the vectors is weighed with $\varpi$.
The black dashed line shows the Alfv\'en surface.
}\label{FRun_mag_strong_pnewBB-103}
\end{figure}

Figures~\protect\ref{FRun_mag_strong-pangmom-98} and \ref{FRun_mag_strong-pangmom-102}
show the distribution of specific angular momentum for the two cases, where
the star is connected to and disconnected from the disc, respectively.
Comparing these two figures,
one can see that in periods of no net accretion, when the star is disconnected
from the disc, specific angular momentum transport by the inner disc wind
is visibly enhanced, carried by enhanced stellar and inner disc winds.
Their velocities are larger in these periods:
terminal outflow speeds are then about $20 \km\s^{-1}$ in the stellar wind
and about $400 \km\s^{-1}$ in the conical shell of the disc wind, where
velocities are highest.
Accretion flow velocities in the disc are up to about $15 \km\s^{-1}$.
In our computational domain,
both the inner stellar wind and the funnel flow remain subsonic
at all times. The slow stellar wind is likely to be due
to our boundary conditions that we impose in our magnetospheric models by
fixing the poloidal velocity to be zero in the star.

A peculiar feature of all our models with a disc as cool as in the models
of this paper, is the presence of at least one radial polarity reversal
of the dynamo-generated magnetic field in the disc
(cf.\ Figs~\protect 23-25 of Paper~I).
In the model with strong magnetosphere discussed in this section,
the first reversal occurs roughly at the corotation radius
($\varpi_{\rm co}\approx 1$), where the angular velocity is about
$7.5 \times 10^{-6}\s^{-1}$, corresponding to a rotation period of around
10~days.

The results in Sects~\ref{MM} and \ref{SM} concerning the oscillating
magnetosphere and associated reconnection processes and episodic accretion,
look indeed like being in the regime described by Goodson \& Winglee (1999).
However, our stellar wind is relatively slow and mostly pressure-driven
whereas their stellar wind (jet) is fast and magneto-centrifugally driven.
On the other hand, the main new feature induced by our disc dynamo is the
structure in our disc wind. Whereas the outer disc wind is slower and mostly
pressure-driven, the inner disc wind is fast, and launched and accelerated
magneto-centrifugally; it might collimate at larger heights to form the
observed protostellar jet. The magnetospheric extension is limited by the
dynamo-generated disc field that is strongest in the inner disc parts and
advected into the disc corona.

\subsection{Dependence on stellar field strength}
\label{Dependence}
\begin{table}[htb]\caption{
Summary of parameter values related to the physics of both the magnetosphere and the disc dynamo
for a sequence of models where the strength of the stellar field is varied.
The hyphen in the row for M1-0 indicates that there is no turbulence in that run.
Note that $|B_{\rm surf}| = A_{\rm star}/r_* \times [\vec{\BB}]$.
}\vspace{12pt}\centerline{\begin{tabular}{lccccccc}
Model & $A_{\rm star}$ & $|B_{\rm surf}| [\kG]$ & $\alpha_0$\\
\hline
N    &  0 & 0   &$-0.15$\\
W    &  1 & 0.2 &$-0.1 $\\
M1   &  5 & 1   &$-0.1 $\\
M1-0 &  5 & 1   &$ 0   $\\
M2   & 10 & 2   &$-0.1 $\\
S    & 25 & 5   &$-0.1 $\\
\label{Tmodels}\end{tabular}}\end{table}
In Table~\ref{Tmodels} we summarize the parameters that have been
changed in the different models considered in this paper and that are
related to the physics of both the magnetosphere and the disc dynamo.
The strength of the stellar field is given both in nondimensional and
in dimensional units, together with the $\alpha_0$ coefficient
quantifying the strength of the $\alpha$ effect
($\alpha_0<0$ means that $\alpha$ is negative in the upper disc plane).

As the strength of the stellar field is reduced, the size of the stellar
magnetosphere shrinks.
At the same time, the field in the disc also decreases somewhat.
Although the conical structure becomes more pronounced as the strength of the stellar
field is increased,
the overall magnetic field and outflow structures remain the same
for medium and strong stellar surface magnetospheric field strengths.
In particular, the field structure (inside the corotation radius) is not of
{\sf X}-point topology, in contrast to what is described by
Shu et al.\ (1994).
Instead, the dynamo-generated magnetic field in the disc always arranges itself
such that, in the corona, the field threading the disc is anti-aligned with the
stellar dipole. This necessarily leads to magnetospheric current sheets; see
Fig.~\protect\ref{FRun_mag_strong_pnewBB-103}.

\subsection{A model with no magnetosphere (Model~N)} \label{nomag}

\begin{figure}[t!]
\centerline{
   \includegraphics[width=8.5cm]{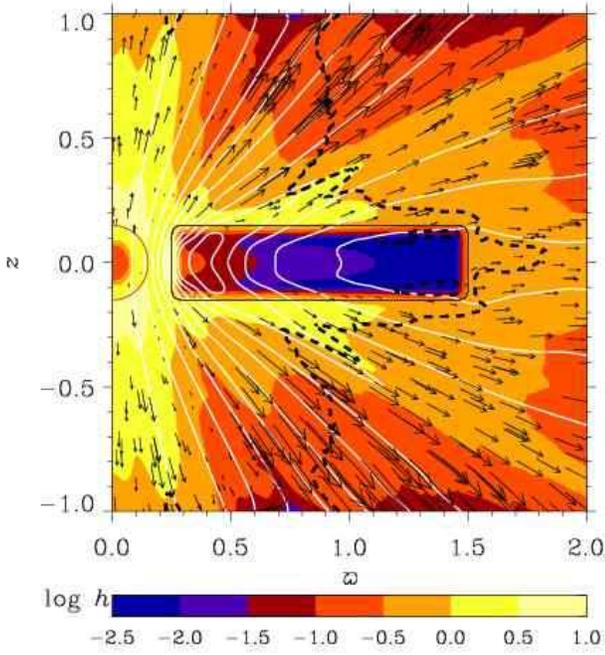}
}
\caption[]{
Similar to Fig.~\protect\ref{FRun_mag_medium1-pnew-1267}, but for the
same model as in Fig.~25 of Paper~I, where now we have averaged over later times
$t=584 \dots t=590$, when the disc dynamo is saturated.
This corresponds to Model~N (no magnetosphere and disc dynamo)
in this paper.
(In this figure, the black dashed line shows the surface where
the poloidal velocity equals $(\cs^2+v_{\rm A,pol}^2)^{1/2}$,
with $\cs$ the sound speed and $v_{\rm A,pol}$ the Alfv\'en speed
from the poloidal magnetic field.) Note the similarity of the overall
outflow and coronal magnetic field structures to Model~S;
Fig.~\protect\ref{FRun_mag_strong-pnew-103}.
}\label{FRun_cold_st}
\end{figure}

In this section we present a model without stellar magnetosphere
where the only source of magnetic fields is the dynamo operating in the disc.
The field threading the star results entirely from advection of the
dynamo-generated disc field and is therefore maintained by the disc dynamo.
In this model without magnetosphere, there are no restrictions to the magnetic
field in the star.
Figure~\protect\ref{FRun_cold_st} shows a time-averaged picture of the same model
as in Fig.~25 of Paper~I, but at later times, when the disc dynamo is saturated.
Note that in this model the inner and outer disc radii are smaller than
in the models with magnetosphere.

\begin{figure}[t!]
\centerline{
   \includegraphics[width=8.5cm]{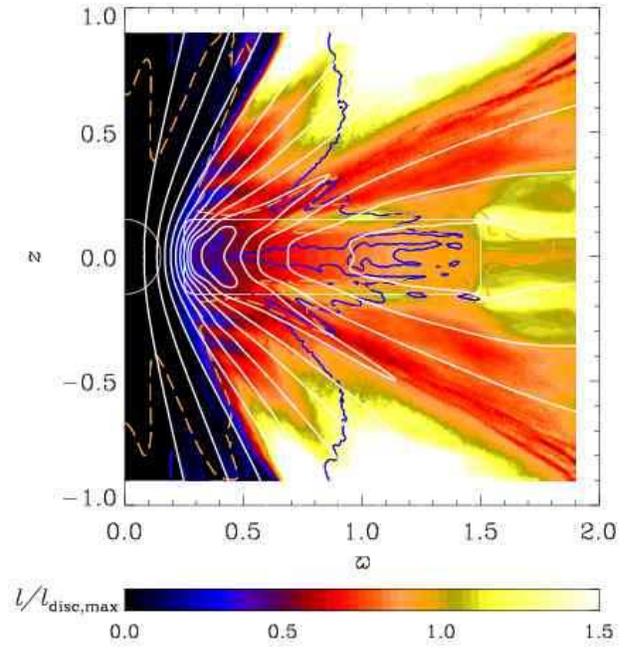}
}
\caption[]{
Shown is the same as in Figs~\protect\ref{FRun_mag_strong-pangmom-98} and
\ref{FRun_mag_strong-pangmom-102}, but for the
same model (Model~N) and time as in Fig.~\protect\ref{FRun_cold_st}.
}\label{FRun_cold_st_pangmom_B}
\end{figure}

The overall outflow and coronal magnetic field structures are very similar
to Model~S (Fig.~\protect\ref{FRun_mag_strong-pnew-103}),
as are the driving mechanisms,
cf.\ Fig.~\protect\ref{FRun_cold_st_pangmom_B} with
Figs~\protect\ref{FRun_mag_strong-pangmom-98} and
\ref{FRun_mag_strong-pangmom-102}.
However, as one can see in Figs~\protect\ref{FRun_cold_st} and
\ref{FRun_cold_st_pangmom_B},
here the stellar wind is much faster and becomes supersonic;
in contrast to Model~S, in this model
without magnetosphere the velocity can freely evolve in the star,
and the only prescription for the star is to
act as a self-regulatory mass sink (cf.\ Sect.~\ref{BE}).
The terminal stellar wind velocity is about $150 \km \s^{-1}$, and
the terminal inner disc wind velocity in the conical shell is also higher
(about $500 \km \s^{-1}$).

Accretion is not episodic as in the magnetospheric models but
the accretion rate is always positive (around $10^{-6}M_\odot\yr^{-1}$),
and the accretion flow is in the cylindrical radial direction. The disc wind
mass loss rate is around $4 \times 10^{-7}M_\odot\yr^{-1}$.
Both rates are therefore higher than the rates in the models with
magnetosphere. The disc wind mass loss rate is only about twice as large,
but the accretion rate is at least $50$ times larger,
compared to when
the episodic accretion in the magnetospheric models reaches its maximum rate.
This changes the ratio between mass accretion rate and disc wind mass loss rate
not only quantitatively, but also qualitatively.
This ratio is now
about $7:3$, but it was less than its reciprocal value (about $1:9$ at most) in the models
with magnetosphere.

Nevertheless, the fact that the stellar magnetic field is still
able to shield a relatively large part
of the accretion flow and deflect it into the wind could be the result
of the assumption of fully axisymmetric flows.
Actual T~Tauri stars possess highly nonaxisymmetric, non-dipolar fields
(Johns-Krull et al.\ 1999, Johns-Krull \& Valenti 2001).
However, the ratio of accretion rate to wind mass loss rate is difficult
to determine observationally with great confidence.
Although Pelletier \& Pudritz (1992) estimate that only about 10\%
of the matter joins the wind,
our ratios lie still in the range of observationally
derived ratios, even those for our models with magnetosphere.

\subsection{Magnetic and accretion torques} \label{MAAT}

In an axisymmetric system,
the equation of conservation of angular momentum
reads
\EQ
{\partial \over \partial t} (\varrho l)
=-\vec{\nabla}\cdot (\vec{t}_{\rm acc} + \vec{t}_{\rm mag} + \vec{t}_{\rm visc}),
\label{AngMomCons}
\EN
where
$l = \varpi u_\varphi = \varpi^2 \Omega$ is the specific angular momentum,
\EQ
\vec{t}_{\rm acc} = \varrho \varpi \uu_{\rm pol} u_\varphi
= \varrho \uu_{\rm pol} \varpi^2 \Omega
\EN
is the material stress
($\uu_{\rm pol}$ is the poloidal velocity field),
\EQ
\vec{t}_{\rm mag} = -\varpi {\BB_{\rm pol} B_\varphi \over 4\pi}
\EN
is the magnetic stress
($\BB_{\rm pol}$ is the poloidal magnetic field), and
\EQ
\vec{t}_{\rm visc} = -\varrho \nu_{\rm t} \varpi^2 \vec{\nabla} \Omega
\EN
is the viscous stress ($\nu_{\rm t}$ is the turbulent kinematic viscosity).
At any given time, the total torque $T$ acting on a central object (star)
is given by the volume integral
\EQ
T = -\int \vec{\nabla}\cdot
    (\vec{t}_{\rm acc} + \vec{t}_{\rm mag} + \vec{t}_{\rm visc}) \, {\rm d}V,
\EN
where the integral has to be taken over a volume enclosing the star.
This is equivalent to the surface integral
\begin{eqnarray}
T &=& -\oint (\vec{t}_{\rm acc} + \vec{t}_{\rm mag} + \vec{t}_{\rm visc})
             \cdot {\rm d}\vec{S} \\
  &=& \oint (-\varrho \uu_{\rm pol} \varpi^2 \Omega
             + \varpi {\BB_{\rm pol} B_\varphi \over 4\pi}
             + \varrho \nu_{\rm t} \varpi^2 \vec{\nabla} \Omega)
             \cdot {\rm d}\vec{S},
\end{eqnarray}
where ${\rm d}\vec{S}$ is the outward directed surface element.
In other words, the total torque acting on a star
is given by the angular momentum flux
across a surface enclosing the star (see Ghosh \& Lamb 1979b).
If the flux is negative, i.e.\ towards the star,
then the torque on the star is positive.

\begin{figure}[t!]
\centerline{
   \includegraphics[width=9.0cm]{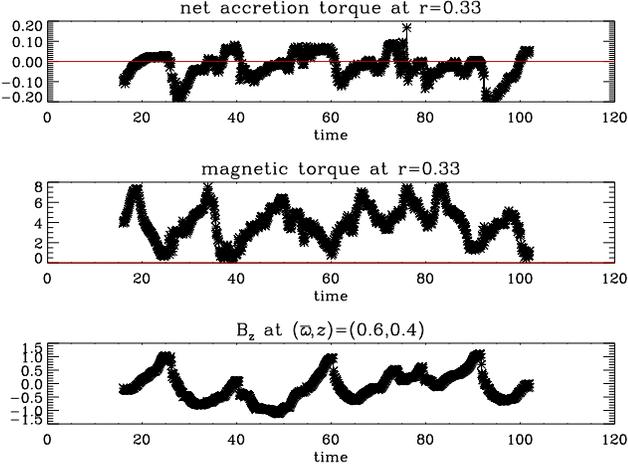}
}
\caption[]{
Dependence of the net accretion torque $T_{\rm acc,net}$ (upper panel) and
the magnetic torque $T_{\rm mag}$ (middle panel) on time
at the spherical radius $r \approx 0.33$ (close to the star) for Model~S.
Lower panel: same indicator for the magnetospheric oscillations as in
Fig.~\protect\ref{pfaveracc_star-two_Bz}, lower panel, and
Fig.~\protect\ref{pfaveracc_disc-two_Bz}, lower panel.
}\label{pptimestorques}
\end{figure}

We calculate the angular momentum flux
towards the star
-- carried by the matter ($T_{\rm acc}$),
by the magnetic field ($T_{\rm mag}$), and due to the viscous stress
($T_{\rm visc}$) -- across spheres around the star, so that we can write
in spherical polar coordinates $(r,\Theta,\varphi)$ (with $\Theta$
the co-latitude, $\varpi=r\sin\Theta$ and $\vec{e}_r$ the radial unit vector):
\begin{eqnarray}
T_{\rm acc}(r) &=& -2\pi r^2\int_0^\pi
\left(\vec{t}_{\rm acc} \cdot \vec{e}_r\right) \sin\Theta \, {\rm d}\Theta \\
&=& -2\pi r^2\int_0^\pi \varrho u_r \varpi^2 \Omega \sin\Theta \, {\rm d}\Theta
\end{eqnarray}
for the accretion torque,
\begin{eqnarray}
T_{\rm mag}(r) &=& -2\pi r^2\int_0^\pi
\left(\vec{t}_{\rm mag} \cdot \vec{e}_r\right) \sin\Theta \, {\rm d}\Theta \\
&=&  2\pi r^2\int_0^\pi \varpi {B_r B_\varphi \over 4\pi} \sin\Theta \, {\rm d}\Theta
\end{eqnarray}
for the magnetic torque, and
\begin{eqnarray}
T_{\rm visc}(r) &=& -2\pi r^2\int_0^\pi
\left(\vec{t}_{\rm visc} \cdot \vec{e}_r\right) \sin\Theta \, {\rm d}\Theta \\
&=& 2\pi r^2\int_0^\pi \varrho \nu_{\rm t} \varpi^2 {\partial \Omega \over \partial r}
    \sin\Theta \, {\rm d}\Theta
\end{eqnarray}
for the viscous torque.
In our units, $T_{\rm [acc,mag,visc]}=1$ corresponds to about
$2.5\times 10^{38}\erg$.

\begin{figure}[t!]
\centerline{
   \includegraphics[width=9.5cm]{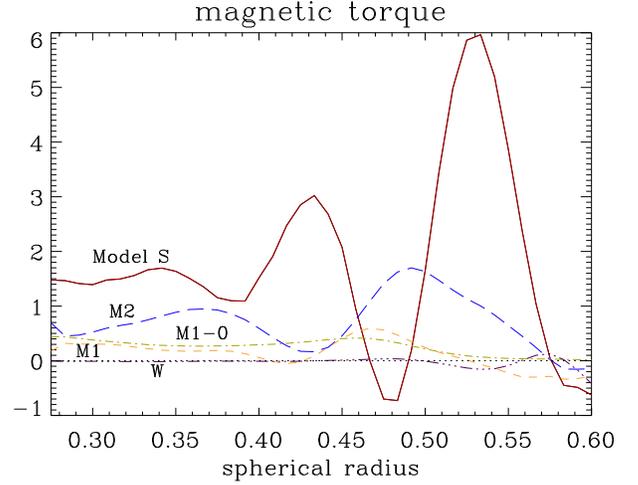}
}
\caption[]{
Magnetic torque $T_{\rm mag}$ as function of spherical radius $r$
for the magnetospheric models presented in this paper. $T_{\rm mag}$ is shown
for radii between the approximate anchoring radius $r_0 = 0.275$
and the inner disc radius $\varpi = 0.6$.
Purple dash-triple-dotted: Model~W, averaged over times $t= 825 \dots t= 850$;
orange dashed:       Model~M1,   $t=1267$;
green dash-dotted: Model~M1-0, averaged over times $t=1100 \dots t=1127$;
blue long-dashed:    Model~M2,   $t= 562$;
red solid:           Model~S,    $t= 103$.
The black dotted line marks the magnetic torqueless state ($T_{\rm mag} = 0$).
}\label{pltorques_runs}
\end{figure}

\begin{figure}[t!]
\centerline{
   \includegraphics[width=9.5cm]{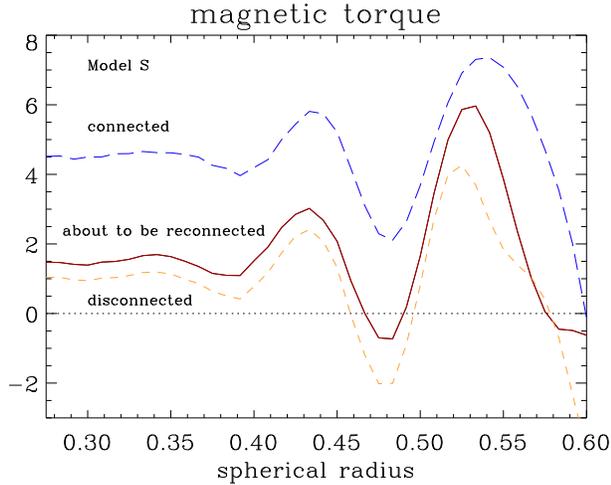}
}
\caption[]{
Magnetic torque $T_{\rm mag}$ as function of spherical radius $r$
for Model~S at three different times. $T_{\rm mag}$ is shown
for radii between the approximate anchoring radius $r_0 = 0.275$
and the inner disc radius $\varpi = 0.6$.
Blue long-dashed: $t=98$ when the star is magnetically connected to the disc;
orange dashed: $t=102$ when the star is magnetically disconnected from the disc;
red solid: $t=103$ when the star is about to be magnetically reconnected
to the disc (cf.\ Fig.~\protect\ref{pltorques_runs}).
The black dotted line marks the magnetic torqueless state ($T_{\rm mag} = 0$).
}\label{pBtorque_98_102}
\end{figure}

\begin{figure*}[t!]
\centerline{
   \includegraphics[width=9.0cm]{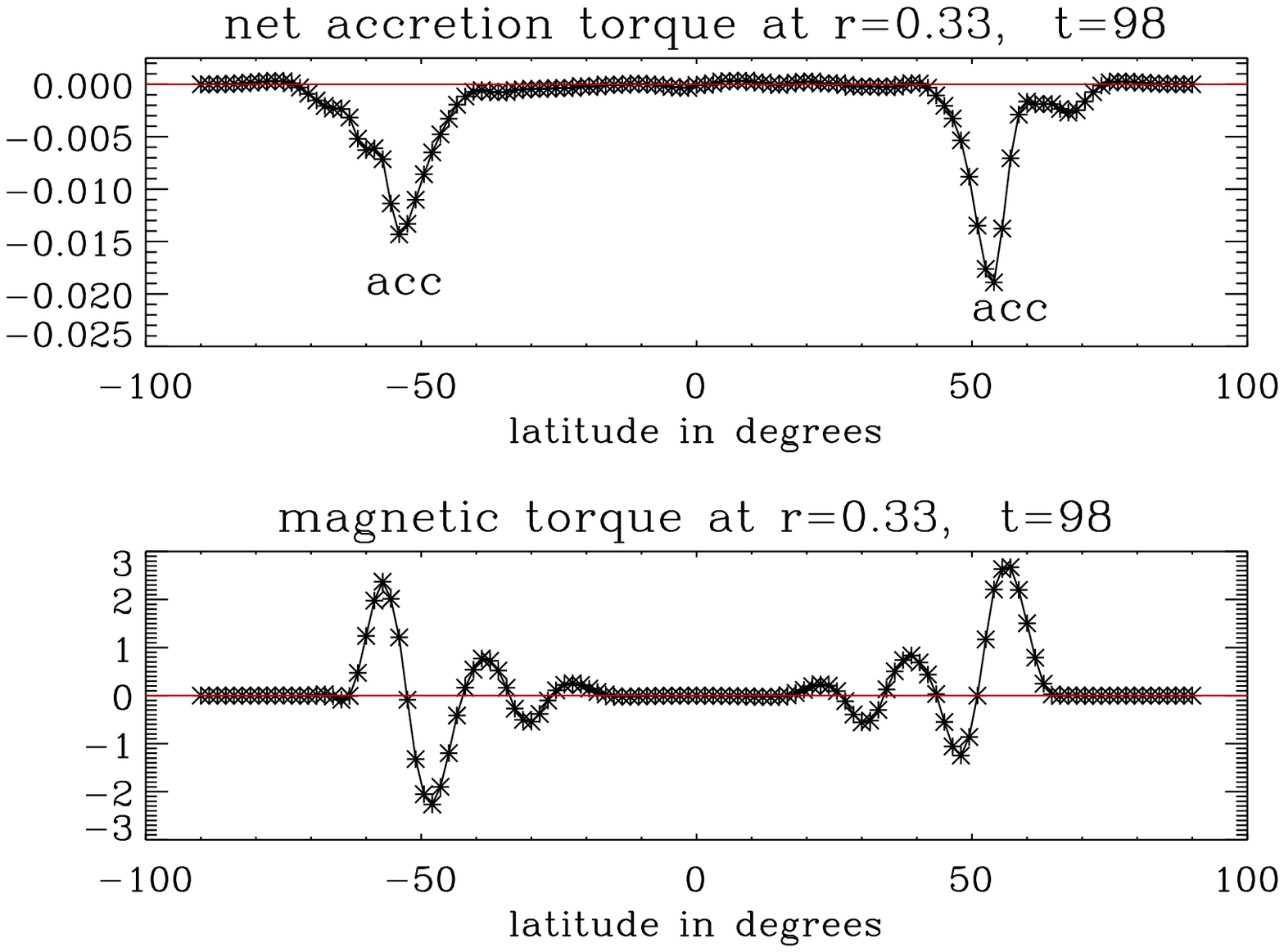}
   \includegraphics[width=9.0cm]{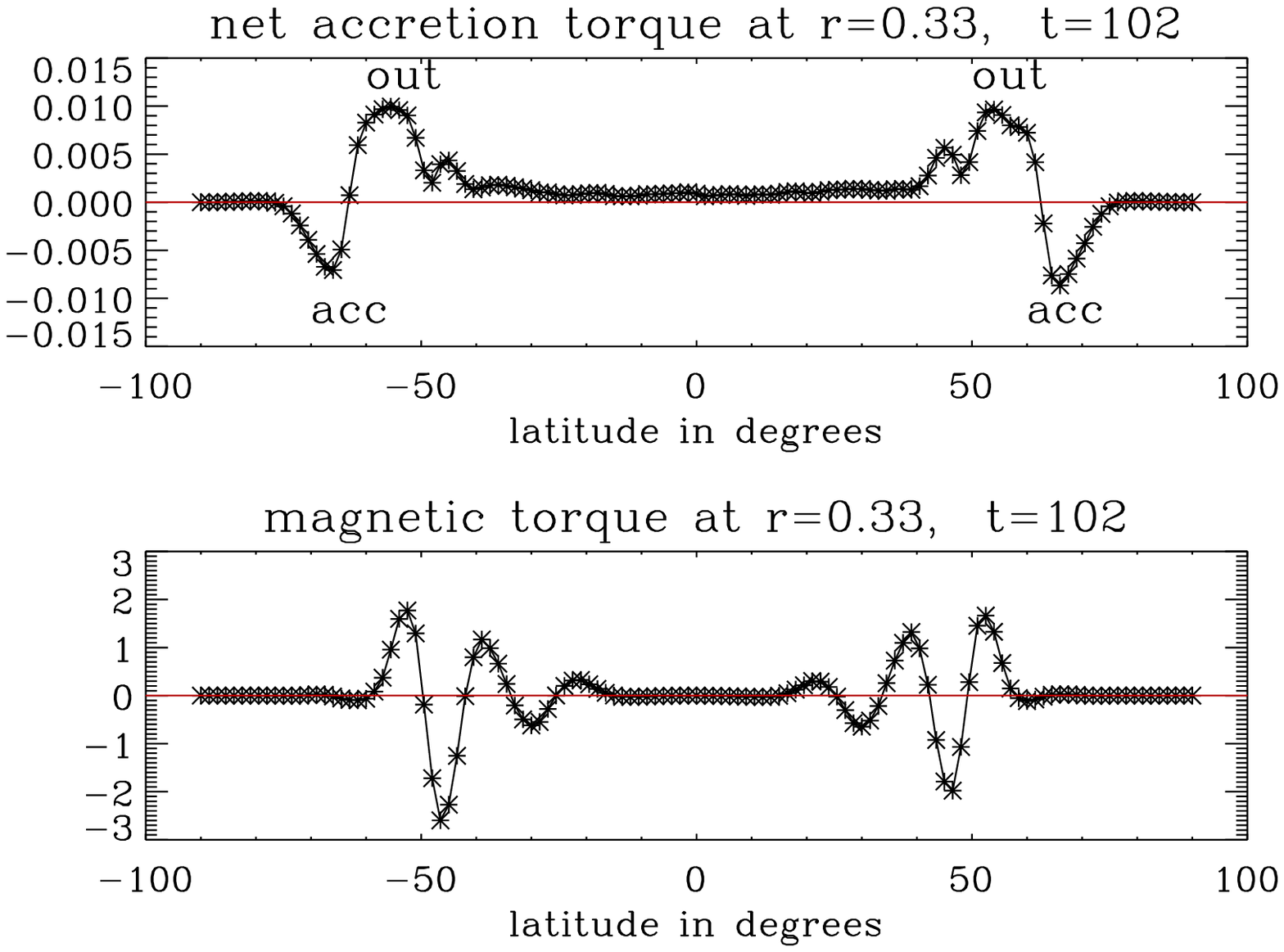}
}
\caption[]{
Net accretion torque (upper panel) and
magnetic torque (lower panel) as functions of latitude
at the spherical radius $r \approx 0.33$ (close to the star) for Model~S
for the two distinctive
states: when the star is magnetically connected to the disc ($t=98$, left)
and when the star is magnetically disconnected from the disc ($t=102$, right).
Latitudes marked with ``acc'' are regions of strongest mass accretion
(cf.\ Fig.~\protect\ref{FRun_mag_strong-p-98}) and
latitudes marked with ``out'' are regions of strongest mass outflow
(cf.\ Fig.~\protect\ref{FRun_mag_strong-p-102}).
}\label{ppThetatorques}
\end{figure*}

At each position between the star and the disc where magnetic field lines
are connected to the star, $B_r B_\varphi > 0$ means
that those field lines are leading the star
in a rotational sense, therefore spinning the star
up. Therefore, $T_{\rm mag} > 0$ translates into a magnetic
spin-up of the star.
Conversely, $B_r B_\varphi < 0$ means
that those field lines are
lagging behind the star and are therefore spinning the star
down, so $T_{\rm mag} < 0$ translates into a magnetic
spin-down of the star.

However, $-u_r \varpi \Omega > 0$ means only that angular
momentum is added to the star. Whether or not this also means a
stellar spin-up (or spin-down if $-u_r \varpi \Omega < 0$),
depends on the magnitude of the angular velocity
of the matter, $\Omega(r,\Theta)$, compared to the magnitude of the
{\it effective} rotation rate
of the stellar surface, $\Omega_*$ (see Appendix~\ref{App} for a
precise definition). We therefore define the net accretion torque,
\begin{equation}
T_{\rm acc,net}(r) = -2\pi r^2\int_0^\pi \varrho u_r \varpi^2 (\Omega-\Omega_*)
\sin\Theta \, {\rm d}\Theta.
\label{NetAccretionTorque}
\end{equation}
If the accreting matter ($u_r<0$) is rotating faster than the
star, the star will be spun up and $T_{\rm acc,net} > 0$.
Likewise, if the accreting matter is rotating slower than the
star, the star will be spun down and $T_{\rm acc,net} < 0$.
On the other hand, a stellar wind ($u_r>0$) that is rotating faster
than the star, will spin down the star, and $T_{\rm acc,net} < 0$,
while a stellar wind that is rotating slower
than the star, will spin up the star, so $T_{\rm acc,net} > 0$.

Figure~\protect\ref{pptimestorques} shows that in Model~S,
close to the star where magnetic field lines are connected to the star,
the net accretion torque is fluctuating around zero, whereas the magnetic torque
is positive at all times. The magnetic torque is always much larger than
the sum of the net accretion and viscous torques ($T_{\rm visc}$
is not shown), suggesting a total spin-up of the star, the angular momentum flux
towards the star
being mainly carried by the magnetic field. A correlation is clearly visible
between $T_{\rm mag}$ and the magnetospheric oscillations, and therefore also
between $T_{\rm mag}$ and the mass accretion rate. When the star is connected
to the disc by its magnetosphere, i.e.\ when $B_z$ is lowest, both
the accretion rate and the magnetic spin-up torque are maximum.

The peak values of the magnetic spin-up torque are around $2\times 10^{39}\erg$,
while those of the material spin-up torque are around $2.5\times 10^{37}\erg$,
and those of the material spin-down torque are around $5\times 10^{37}\erg$.

Also Romanova et al.\ (2002) find that most of the angular momentum flux
to the star is carried by the magnetosphere. In the case of a protostar
rotating with a period of around 9.4~days and a stellar surface field strength
of about $1.1\kG$, they find that matter carries only about $1\%$ of the total
flux. The magnetic torque is positive so that it acts to spin up the star,
and its amplitude is also correlated with the accretion rate.

In all our models with sufficiently strong magnetosphere
($|B_{\rm surf}|\ge 1 \kG$;
Models~M1, M1-0, M2 and S), $T_{\rm mag}$ is mainly positive
between the anchoring region and the inner disc edge ($0.275 < \varpi < 0.6$);
see Fig.~\protect\ref{pltorques_runs}.
This is roughly the region where magnetic lines are connected to the star,
so that the star experiences the torque. This indicates that for sufficiently strong
magnetosphere, angular momentum will be added to the star, carried
by the magnetic field, resulting in a stellar spin-up by the magnetic field.
As the stellar surface field strength of the magnetosphere increases,
also the magnetic torque at radii between star and disc generally increases.
In this region, the accretion and viscous torques (not shown) are negligible
compared to the magnetic torque in all our magnetospheric models.

Figure~\protect\ref{pBtorque_98_102} shows that when star and disc
are magnetically connected, the magnetic spin-up torque is higher
at all spherical radii between
the approximate anchoring radius $r_0 = 0.275$ and the inner disc radius
$\varpi = 0.6$.
Basically what happens is this:
when the star is connected to the disc, there is accretion along magnetospheric
field lines, and a large positive magnetic torque leads to a magnetic spin-up
of the star. Then the star disconnects from the disc, and the inner disc edge
is magnetically spun up (see Fig.~\protect\ref{pBtorque_98_102});
an enhanced inner disc wind carries away excess specific angular
momentum (see also Fig.~\protect\ref{FRun_mag_strong-pangmom-102}).

The latitudinal dependencies of the torques for Model~S, shown in
Fig.~\protect\ref{ppThetatorques}, reveal that there is also a correlation
between the net accretion torque [i.e.\ the non-integrated quantity of
Eq.~(\ref{NetAccretionTorque})] and the mass accretion rate. At latitudes of
accretion onto the star, the net accretion torque is negative
(i.e.\ a material spin-{\it down}), whereas at latitudes of stellar outflow,
the net accretion torque is positive (i.e.\ a material spin-{\it up}).
This is due to the small rotation rates between the star and the disc.

\section{Conclusions}

In agreement with earlier work by Hirose et al.\ (1997),
Goodson \& Winglee (1999) and
Matt et al.\ (2002), our work confirms the possibility of episodic
and recurrent magnetospheric accretion also for models where the disc magnetic
field is not imposed but dynamo-generated.
The critical stellar surface field strength required for the episodic
accretion to be correlated with the magnetic star--disc coupling,
is around (or below)
$2\kG$, which is well in the range of observed field strengths for
T~Tauri stars (Johns-Krull et al.\ 1999).
These recurrent changes in the connectivity between the stellar magnetosphere and
the dynamo-generated disc field result in episodic mass transfer
from the disc to the star.
For the same stellar surface field strengths, the time-dependent wind
velocities are also correlated with the magnetic star--disc coupling.
The wind velocities as well as specific angular momentum transport
from the disc inner edge into the corona
are enhanced during periods when the star is disconnected from the disc.
Highly time-dependent accretion and outflows have also been detected
observationally in classical T~Tauri stars (CTTS) by Stempels \& Piskunov
(2002, 2003).

The observed stellar fields are, however, not dipolar, but show a
strong nonaxisymmetric component, although the geometry of the magnetosphere
of protostars is yet unknown. [The assumption of a stellar dipolar field
is motivated by observations that suggest that the stellar field
might be concentrated at the poles in rapidly rotating CTTS;
e.g.,\ Sch\"ussler et al.\ (1996)].
The fact that the fields of T~Tauri stars are not axisymmetric
may have important implications for the accretion process.
Our work suggests that dipolar fields tend to divert a significant
fraction of disc matter into the wind.
Channelling the disc material along magnetospheric stellar field lines
becomes more
efficient if the magnetospheric accretion process happens in an episodic
fashion.
It is conceivable that a nonaxisymmetric stellar field has a similar
effect, but it would then produce time variability on a time scale
similar to the stellar rotation period.
Variability on time scales similar to the stellar orbital period
has indeed been observed; see Johns-Krull et al.\ (1999).

We find that the sum of magnetic, accretion and viscous torques
acting on the star,
is positive at most radii between the star and the disc
if the stellar surface field strength exceeds a certain value
that is somewhere between $200\G$ and $1\kG$.
This positive torque means a stellar spin-up, in agreement with
Romanova et al.\ (2002) for stellar rotation periods of around 9 to 10~days.
The accretion and viscous torques are almost always much smaller
(in amplitude) than the magnetic torque. The small accretion torque
in our simulations
might be due to the relatively small accretion rates and accretion flow
velocities, both of which are not imposed.
These, in turn, might be due to the fixed inner disc radius.

The mean-field disc dynamo is responsible for the structure in the disc wind,
with coexisting pressure-driving of the outer disc wind and magneto-centrifugal
launching and acceleration of the inner disc wind. The stellar wind
is always mostly pressure-driven.

In contrast to the simulations of Hirose et al.\ (1997) and
Goodson et al.\ (1997), we do not find any stellar (fast) jet
that is driven by magneto-centrifugal processes. However, our
magneto-centrifugally driven fast inner disc wind is close to the star
and might collimate at larger heights to form the observed protostellar jet
(cf.\ Fendt et al. 1995).

Another important result that has emerged from the present investigations
is an anti-alignment of the disc magnetic field relative to the stellar
dipole (no {\sf X}-point).
In our models, the relative orientation of the disc field and the stellar field
is no longer a free input parameter, but a result of the simulations.
The only obvious way to prevent anti-alignment of the disc magnetic
field, is to have magnetic fields in the disc and in the protostar that are
entirely due to the accretion of an ambient large scale field.
This may indeed be quite plausible for many protostellar discs in a large
fraction of star forming regions.
Another argument in favour of this possibility is the fact that strong
collimation of protostellar outflows into well pronounced jets has so far
only been found in the presence of a large scale field aligned with
the rotation axis of the star-disc system (e.g.,\ Ouyed et al.\ 1997), and
not for dynamo-generated disc magnetic fields (see Paper~I).

An important limitation of the present work is the restriction to
axisymmetric magnetospheric accretion.
Given that fully three-dimensional simulations have now become feasible
(e.g.,\ Hawley 2000), this would certainly be an important constraint to
be relaxed in future simulations.

\appendix
\section{Effective stellar rotation rate}
\label{App}

In this appendix we give a precise definition of the
{\it effective} stellar rotation rate, $\Omega_*$, as it was used
in Eq.~(\ref{NetAccretionTorque}).

Integrating Eq.~(\ref{AngMomCons}) over the stellar volume
yields an evolution equation for the total stellar angular momentum, $L_*$,
\EQ
\dot{L}_*=T_{\rm acc}+T_{\rm mag}+T_{\rm visc},
\label{Ldot}
\EN
where $L_*=\int\varrho l\,\dd V$ is the integral of the angular momentum density,
$\varrho l$, over the
stellar volume.
Obviously, $L_*$ can grow from mass accretion alone
without spinning up the star.
We therefore need to look at the evolution of the specific stellar angular momentum,
$l_*=L_*/M_*$, where $M_*=\int\varrho\,\dd V$ is the mass of the star.
Using the product rule, we can write the left hand side of Eq.~(\ref{Ldot}) as
\EQ
\dot{L}_*=\dot{l}_*M_*+l_*\dot{M}_*,
\EN
where the accretion rate $\dot{M}_*$ can be expressed in terms of the
mass flux density, $\varrho\uu_{\rm pol}$, via $\dot{M}_*=-\oint\varrho\uu_{\rm pol}\cdot\dd\SSS$,
where the integral is taken over the stellar surface ($\dd\SSS$ is the outward directed
surface element).
Further, $l_*=L_*/M_*=\int\varrho l\,\dd V / \int\varrho\,\dd V$ can be expressed as
$l_*=\varpi_*^2\Omega_*$, where $\varpi_*$ is defined as a weighed average,
\EQ
\varpi_*^2\equiv
\langle\varpi^2\rangle=\left.
\oint\varrho\uu_{\rm pol}\varpi^2\cdot\dd\SSS
\right/\oint\varrho\uu_{\rm pol}\cdot\dd\SSS,
\EN
where the integrals are taken over the stellar surface.
This defines $\Omega_*$,
which replaces the intuitive definition of $\Omega_*$ as the
stellar surface rotation rate (the latter, however, is still sufficiently
accurate a definition for all practical purposes).

With this definition of $\varpi_*$ (and $\Omega_*$),
we obtain an evolution equation for the specific stellar angular momentum, $l_*$,
\EQ
M_*\dot{l}_*=T_{\rm acc}-l_*\dot{M}_*+T_{\rm mag}+T_{\rm visc},
\EN
which can be written as
\EQ
M_*\dot{l}_*=T_{\rm acc,net}+T_{\rm mag}+T_{\rm visc}
\label{Mldot}
\EN
with $T_{\rm acc,net}$ given in Eq.~(\ref{NetAccretionTorque}).
This shows that the sign of the right hand side of Eq.~(\ref{Mldot})
determines whether the star spins up or down.

\begin{acknowledgements}
Use of the supercomputer SGI 3800 in Link\"oping and
of the PPARC supported supercomputers in St Andrews and Leicester is acknowledged.
This research was conducted using the resources of
High Performance Computing Center North (HPC2N).
B.v.R.\ thanks NORDITA for hospitality.
We thank Eric Blackman, Sean Matt and Ulf Torkelsson for fruitful
discussions.
We also thank an anonymous referee for many useful comments.
\end{acknowledgements}

\newcommand{\yapj}[3]{ #1, {ApJ,} {#2}, #3}
\newcommand{\ypasj}[3]{ #1, {Publ. Astron. Soc. Japan,} {#2}, #3}
\newcommand{\yana}[3]{ #1, {A\&A,} {#2}, #3}
\newcommand{\ymn}[3]{ #1, {MNRAS,} {#2}, #3}
\newcommand{\ybook}[3]{ #1, {#2} (#3)}
\newcommand{\ynat}[3]{ #1, {Nat,} {#2}, #3}
\newcommand{\ypr}[3]{ #1, {Phys. Rev.} {#2}, #3}
\newcommand{\yjgr}[3]{ #1, {JGR,} {#2}, #3}
\newcommand{\yproc}[5]{ #1, in {#3}, ed. #4 (#5), #2}

\vfill\bigskip\noindent\tiny\it{


\begin{thebibliography}{}

\bibitem{AP00}
Agapitou, V., \& Papaloizou, J.~C.~B.\ymn{2000}{317}{273}

\bibitem{Bar99}
Bardou, A.\ymn{1999}{306}{669}

\bibitem{RekowskiEtal}
Bardou, A., von Rekowski, B., Dobler, W., et al. 2001, A\&A, 370, 635

\bibitem{BlandfordPayne}
Blandford, R.~D., \& Payne, D.~R. 1982, MNRAS, 199, 883

\bibitem{Brandenb98}
Brandenburg, A.\yproc{1998}{61}
{Theory of Black Hole Accretion Discs}
{M. A. Abramowicz, G. Bj\"ornsson \& J. E. Pringle}
{Cambridge University Press}

\bibitem{BrandenbNordlundEtal}
Brandenburg, A., Nordlund, \AA, Stein, R.~F., et al. 1995, ApJ, 446, 741

\bibitem{BrandenbTuominKrause}
Brandenburg, A., Tuominen, I., \& Krause, F. 1990, GAFD, 50, 95

\bibitem{Campbell}
Cameron, A.~C., \& Campbell, C.~G.\yana{1993}{274}{309}

\bibitem{Campbell99}
Campbell, C.~G. 1999, MNRAS, 310, 1175

\bibitem{Campbell00}
Campbell, C.~G. 2000, MNRAS, 317, 501

\bibitem{Campbell01}
Campbell, C.~G. 2001, MNRAS, 323, 211

\bibitem{Dobler_etal02}
Dobler, W., Shukurov, A., \& Brandenburg, A.\ypr{2002}{E 65}{036311}

\bibitem{Fendt_etal95}
Fendt, C., Camenzind, M., \& Appl, S. 1995, A\&A, 300, 791

\bibitem{GL79a}
Ghosh, P., \& Lamb, F.~K.\yapj{1979a}{232}{259}

\bibitem{GL79b}
Ghosh, P., \& Lamb, F.~K.\yapj{1979b}{234}{296}

\bibitem{GLP77}
Ghosh, P., Lamb, F.~K., \& Pethick, C.~J. \yapj{1977}{217}{578}

\bibitem{GBW99}
Goodson, A.~P., B\"ohm, K.-H. \& Winglee, R.~M.\yapj{1999}{524}{142}

\bibitem{GW99}
Goodson, A.~P., \& Winglee, R.~M.\yapj{1999}{524}{159}

\bibitem{GWB97}
Goodson, A.~P., \& Winglee, R.~M., B\"ohm, K.-H.\yapj{1997}{489}{199}

\bibitem{GuentherEtal99}
Guenther, E.~W., Lehmann, H., Emerson, J.~P., et al. 1999, A\&A, 341, 768

\bibitem{JanhunenHuuskonen93}
Janhunen, P., \& Huuskonen, A.\yjgr{1993}{98}{9519}

\bibitem{JKVK99}
Johns-Krull, C.~M., \& Valenti, J.~A.\yapj{2001}{561}{1060}

\bibitem{JKV01}
Johns-Krull, C.~M., Valenti, J.~A., \& Koresko, C.\yapj{1999}{516}{900}

\bibitem{Haw00}
Hawley, J.~F.\yapj{2000}{528}{462}

\bibitem{HB91}
Hawley, J.~F., \& Balbus, S.~A.\yapj{1991}{376}{223}

\bibitem{HSM96}
Hayashi, M.~R., Shibata, K., \& Matsumoto, R.\yapj{1996}{468}{L37}

\bibitem{HUSM97}
Hirose, S., Uchida, Y., Shibata, K., et al.\ypasj{1997}{49}{193}

\bibitem{Koe91}
K\"onigl, A.\yapj{1991}{370}{L39}

\bibitem{KrauseRaedler}
Krause, F., \& R\"adler, K.-H. 1980, Mean-Field Magnetohydrodyna{-}mics and
Dynamo Theory, Akademie-Verlag, Berlin

\bibitem{KHR03}
K\"uker, M., Henning, T., \& R\"udiger, G., 2003, ApJ, 589, 397

\bibitem{LRBK95}
Lovelace, R.~V.~E., Romanova, M.~M.,
\& Bisnovatyi-Kogan, G.~S.\yapj{1995}{275}{244}

\bibitem{MGWB02}
Matt, S., Goodson, A.~P., Winglee, R.~M., et al.\yapj{2002}{574}{232}

\bibitem{SM97}
Miller, K.~A., \& Stone, J.~ M.\yapj{1997}{489}{890}

\bibitem{OPS97}
Ouyed, R., Pudritz, R.~E., \& Stone, J.~M.\ynat{1997}{385}{409}

\bibitem{PT94}
Papaloizou, J.~C.~B., \& Terquem, C. 1999, ApJ, 521, 823

\bibitem{Par79}
Parker, E.~N.\ybook{1979}{Cosmical Magnetic Fields}{Clarendon Press, Oxford}

\bibitem{PellePud}
Pelletier, G., \& Pudritz, R.~E. 1992, ApJ, 394, 117

\bibitem{RekowskiEtal2002}
von Rekowski, B., Brandenburg, A., Dobler, W., et al. 2003, A\&A, 398, 825
(Paper~I)

\bibitem{RekowskiEtal2000}
von Rekowski, M., R\"udiger, G., \& Elstner, D. 2000, A\&A, 353, 813

\bibitem{RS95}
Reyes-Ruiz, M., \& Stepinski, T.~F. 1995, ApJ, 438, 750

\bibitem{RUKL02}
Romanova, M.~M., Ustyugova, G.~V., Koldoba, A.~V., et al.\yapj{2002}{578}{420}

\bibitem{Sch96}
Sch\"ussler, M., Caligari, P., Ferriz-Mas, A., et al.\ 1996, A\&A, 314, 503

\bibitem{ShuEtal1994}
Shu, F., Najita, J., Ostriker, E., et al.\yapj{1994}{429}{781}

\bibitem{StempelsPiskunov2002}
Stempels, H.C., \& Piskunov, N. 2002, A\&A, 391, 595

\bibitem{StempelsPiskunov2003}
Stempels, H.C., \& Piskunov, N. 2003, A\&A, 408, 693

\bibitem{Tuominenetal94}
Tuominen, I., Brandenburg, A., Moss, D., et al.\yana{1994}{284}{259}

\bibitem{UB98}
Urpin, V., \& Brandenburg, A.\ymn{1998}{294}{399}

\bibitem{Yousefetal03}
Yousef, T. A., Brandenburg, A., \& R\"udiger, G. 2003, A\&A, 411, 321

\bibitem{ZieglerRuediger}
Ziegler, U., \& R\"udiger, G. 2000, A\&A, 356, 1141

\end{thebibliography}
\end{document}